\documentclass[journal]{IEEEtran}
\usepackage{cite}
\ifCLASSINFOpdf
\else
\fi
\usepackage{amsmath}
\usepackage{algorithmic}
\usepackage{array}
\usepackage{graphicx}
\usepackage{amsfonts}
\usepackage{caption}
\usepackage{algorithm}
\usepackage{subfigure}
\usepackage{bm}
\usepackage{color}
\usepackage{booktabs}

\newtheorem{remark}{Remark}
\newcommand{\tabincell}[2]{}

\begin{document}

\title{Massive Access in Cell-Free Massive MIMO-Based Internet of Things: Cloud Computing and Edge Computing Paradigms}

\author{Malong~Ke,~\IEEEmembership{Student~Member,~IEEE,}
        Zhen~Gao,~\IEEEmembership{Member,~IEEE,}
        Yongpeng~Wu,~\IEEEmembership{Senior~Member,~IEEE,}
        Xiqi~Gao,~\IEEEmembership{Fellow,~IEEE,}
        and~Kat-Kit~Wong,~\IEEEmembership{Fellow,~IEEE}
\thanks{Manuscript received February 1, 2020; revised June 9, 2020; accepted July 17, 2020.}
%%Date of publication July 17, 2020; date of current version July 17, 2020.
%The work of Z.~Gao was supported in part by the National Natural Science Foundation of China under Grant 61701027, in part by Beijing Natural Science Foundation under Grants 4182055 and L182024, in part by Young Elite Scientists Sponsorship Program by CAST, and in part by Talent Innovation Project of BIT.
%The work of Y.~Wu was supported in part by the National Key R\&D Program of China under Grant 2018YFB1801102, in part by JiangXi Key R\&D Program under Grant 20181ACE50028, in part by National Science Foundation under Grant 61701301, in part by Young Elite Scientist Sponsorship Program by CAST, and in part by the open research project of State Key Laboratory of Integrated Services Networks (Xidian University) under Grant ISN20-03.
%The work of X.~Gao was supported by the National Key R\&D Program of China under Grant 2018YFB1801103.
%\emph{(Corresponding author: Zhen Gao.)}}
\thanks{M. Ke and Z. Gao are with the School of Information and Electronics, Beijing Institute of Technology, Beijing 100081, China, and also with the Advanced Research Institute of Multidisciplinary Science, Beijing Institute of Technology, Beijing 100081, China (e-mail: kemalong@bit.edu.cn; gaozhen16@bit.edu.cn).}
\thanks{Y. Wu is with the Department of Electronic Engineering, Shanghai Jiao Tong University, Shanghai 200240, China, and also with the State Key Laboratory of Integrated Services Networks, Xidian University, Xian 710071, China (e-mail: yongpeng.wu@sjtu.edu.cn).}
\thanks{X. Gao is with the National Mobile Communications Research Laboratory, Southeast University, Nanjing 210096, China (e-mail: xqgao@seu.edu.cn).}
\thanks{K.-K. Wong is with the Department of Electronic and Electrical Engineering, University College London, London WC1E 6BT, U.K. (e-mail: kai-kit.wong@ucl.ac.uk).}}

\markboth{IEEE JOURNAL ON SELECTED AREAS IN COMMUNICATIONS,~Vol.~14, No.~8, July~2020}%
{Shell \MakeLowercase{\textit{et al.}}: Bare Demo of IEEEtran.cls for IEEE Journals}

%\IEEEpubid{0000--0000/00\$00.00~\copyright~2020 IEEE}


\maketitle

\begin{abstract}
This paper studies massive access in cell-free massive multi-input multi-output (MIMO)-based Internet of Things and solves the challenging active user detection (AUD) and channel estimation (CE) problems.
For the uplink transmission, we propose an advanced frame structure design to reduce the access latency.
Moreover, by considering the cooperation of all access points (APs), we investigate two processing paradigms at the receiver for massive access: cloud computing and edge computing.
For cloud computing, all APs are connected to a centralized processing unit (CPU), and the signals received at all APs are centrally processed at the CPU.
While for edge computing, the central processing is offloaded to part of APs equipped with distributed processing units, so that the AUD and CE can be performed in a distributed processing strategy.
Furthermore, by leveraging the structured sparsity of the channel matrix, we develop a structured sparsity-based generalized approximated message passing (SS-GAMP) algorithm for reliable joint AUD and CE, where the quantization accuracy of the processed signals is taken into account.
Based on the SS-GAMP algorithm, a successive interference cancellation-based AUD and CE scheme is further developed under two paradigms for reduced access latency.
Simulation results validate the superiority of the proposed approach over the state-of-the-art baseline schemes.
Besides, the results reveal that the edge computing can achieve the similar massive access performance as the cloud computing, and the edge computing is capable of alleviating the burden on CPU, having a faster access response, and supporting more flexible AP cooperation.
\end{abstract}

\vspace{-0.5mm}
\begin{IEEEkeywords}
Massive access, cell-free massive MIMO, cloud computing, edge computing, active user detection, structured sparsity.
\end{IEEEkeywords}

\vspace{-1mm}
\section{Introduction}
\label{Sec:I}

\IEEEPARstart{W}{ith} the advent of the Internet-of-Things (IoT) era, massive machine-type communications (mMTC) have been identified as the indispensable services in future wireless networks~\cite{Chen_JSAC'20},~\cite{Dohler_CM'17}.
Against this background, the future base stations (BSs) are expected to enable massive connectivity with billions of user equipments (UEs).
However, the reliable support of low-latency massive access for mMTC is still challenging in current wireless networks~\cite{Bockelmann_Access'18}.
On the one hand, assigning orthogonal pilot sequences to all potential UEs would be impractical for massive access.
On the other hand, for traditional grant-based random access protocols, the complex signaling information interaction would lead to the extremely high access latency when the number of UEs becomes large~\cite{Hasan_CM'13}.
Fortunately, a key characteristic of mMTC is the sporadic traffic of UEs, i.e., among a large pool of UEs, only a small fraction are active in any given time interval~\cite{Liu_SPM'18}.
Hence, the grant-free random access protocol is recently proposed as a promising alternative, where each active UE transmits its pilots and data to the BS simultaneously without scheduling in advance~\cite{Laya_CST'14}.
In grant-free random access, the BS has to utilize the received pilot signals to detect the active UEs and estimate their channels, which are vital for the subsequent data detection~\cite{Shao_IoTJ'19}.
However, due to the large number of UEs but the limited radio resources for massive access, the active user detection (AUD) has been emerging as a challenging problem~\cite{{Laya_CST'14}, {Liu_SPM'18}, {Shao_IoTJ'19}}.

Moreover, since the power limited IoT UEs are usually distributed in a vast area, multiple BSs should cooperate to offer a better coverage and to save the transmit power of UEs.
Different from the massive access for single-BS scenarios, the multiple BS scenarios pose a new massive access problem known as ``multi-cell massive access"~\cite {Chen_TWC'19} or ``random access for crowded massive MIMO systems"~\cite{Han_TWC'17},~\cite{Bjornson_TWC'17}.
For traditional network architecture, each BS operates independently to perform AUD and channel estimation (CE) for the UEs distributed in its own cell while treating the inter-cell interference as noise~\cite{Chen_TWC'19}.
Consequently, the inter-cell interference is a severely limiting factor for reliable massive access.
Fortunately, the promising cell-free massive MIMO network brings new opportunities to facilitate the massive access, where the massive MIMO BSs are regarded as access points (APs) and deployed in a vast area to serve massive IoT UEs, and these APs are connected to one or multiple processing units for joint signal processing~\cite{Chen_JSAC'20},~\cite{Ngo_TWC'17}.
Since there are no ``cells" or ``cell boundaries", the inter-cell interference can be avoided.
However, the design of an efficient AUD and CE scheme for grant-free massive access in cell-free massive MIMO systems is still an open issue.

\vspace{-1mm}
\subsection{Related Work}
\label{sec:I-A}

Exploiting the sparse UE activity, several compressive sensing (CS)-based approaches have been proposed to detect active UEs for grant-free massive access.
In~\cite{Shim_CL'12}, a CS-based multiuser detection method was suggested, where the concerned AUD was formulated as a sparse signal recovery problem.
But this method only considered the detection in one symbol period.
In typical massive access scenarios, the active UEs generally transmit uplink signals in several successive time slots~\cite{Abebe_CL'16}.
By assuming the UE activity remains unchanged in several adjacent time slots, the authors in~\cite{Wang_CL'16} proposed a structured iterative support detection algorithm to jointly detect the active UEs and the transmitted data, where the structured sparsity pattern observed in multiple time slots was leveraged for improved detection performance.
However, for practical IoT applications, the UEs can randomly access or leave the system, which yields a time-varying UE activity.
On the other hand, although the active UE set (AUS) can be changed over time, the variation would be gradual~\cite{Vaswani_TSP'16}.
This leads to the temporal correlation of UE activity within several successive time slots.
Hence, a dynamic CS-based multi-user detection approach was proposed in~\cite{Wang_CL'16_2}, where the AUS obtained in current time slots was used as the a priori information to estimate AUS in the next time slot.
However, the solution~\cite{Wang_CL'16_2} assumes the availability of the sparsity level, i.e., the number of active UEs, which can be unrealistic.
To overcome this shortcoming, in~\cite{Du_JSAC'17}, the authors developed an efficient prior-information-aided adaptive subspace pursuit algorithm to detect active UEs without the knowledge of the sparsity level.
Furthermore, by leveraging the a priori information of the transmitted signals, the authors of~\cite{Wei_CL'17} proposed an approximate message passing (AMP)-based joint AUD and data detection scheme for further improved performance.

The solutions~\cite{{Shim_CL'12}, {Abebe_CL'16}, {Wang_CL'16}, {Vaswani_TSP'16}, {Wang_CL'16_2}, {Du_JSAC'17}, {Wei_CL'17}} focus on joint AUD and data detection, which assume the availability of perfect channel state information (CSI).
In practice, the channels between the active UEs and the BS should be estimated before the following coherent data detection.
Based on the idea of the orthogonal matching pursuit, the authors of~\cite{Schepker_ISWCS'13} proposed an efficient greedy algorithm to realize joint AUD and CE, where only single-antenna is considered at the BS.
The analysis and numerical results in~\cite{Liu_TSP'18} reveal that the detection error probability of AUD can always be driven to zero by equipping a large-scale antenna array at the BS.
Against this background, an advanced grant-free massive access scheme was developed for multi-antenna systems~\cite{Park_SPAWC'17}, where both sparse UE activity and the sparsity of the delay-domain channel impulse response (CIR) were leveraged for facilitating AUD and CE.
To reduce the computational complexity in the case of a large number of UEs and antennas, a dimension reduction-based joint AUD and CE approach was further proposed in~\cite{Shao_TSP'20}.
The solutions~\cite{{Schepker_ISWCS'13}, {Park_SPAWC'17}, {Shao_TSP'20}} are developed from the CS greedy (non-Bayesian) algorithms to achieve the sparse signal recovery, where the a priori distribution of the channels is not taken into account.
By exploiting the statistical information of the massive access channels based on the Bayesian inference framework, the authors in~\cite{Chen_TSP'18} developed an AMP-based access scheme, which could significantly improve the AUD and CE performance compared to the greedy approaches.
Besides, an expectation propagation-based scheme was proposed in~\cite{Ahn_Tcom'19} for further enhanced performance.
However, the work~\cite{{Chen_TSP'18}, {Ahn_Tcom'19}} assumes that the noise variance and the parameters of the a priori distribution of channels are known in advance.
In~\cite{Ke_TSP'20}, an expectation maximization (EM) algorithm was incorporated into the AMP-based scheme to learn the unknown hyper-parameters.
Meanwhile, the structured sparsity of the massive access channel matrix observed at multiple BS antennas was leveraged to improve AUD performance.
Furthermore, the joint AUD and CE for massive access was further extended to the cloud radio network architecture~\cite{He_TWC'18} and multi-cell massive access scenarios~\cite{Chen_TWC'19}.

\vspace{-1mm}
\subsection{Main Contributions}
\label{Sec:I-B}

In this paper, we investigate grant-free massive access in cell-free massive MIMO-based IoT, where orthogonal frequency division multiplexing (OFDM) technique is employed for uplink transmission.
Specifically, we first propose a frame structure design for massive access and then compare two processing paradigms consisting of cloud computing and edge computing for the practical processing of AUD and CE.
Due to the limited capacity of the backhaul links between APs and processing units, we further consider the quantization of the APs' received signals.
For both paradigms, by exploiting the sporadic traffic of UEs and the angular-domain sparsity of massive MIMO channels, the AUD and CE problems are formulated as two CS problems based on the spatial-domain and angular-domain channel models, respectively.
Subsequently, a structured sparsity-based generalized AMP (SS-GAMP) algorithm is developed for CS recovery, where the quantization of the processed signals is considered.
On this basis, a successive interference cancellation (SIC)-based AUD and CE algorithm is developed for alternately detecting active UEs and estimating their channels.
Our main contributions can be summarized as follows:

\begin{figure*}[t]
    \captionsetup{font={footnotesize}, name = {Fig.}, labelsep = period}
    \centering
    \subfigure[]{\includegraphics[width=0.7\columnwidth]{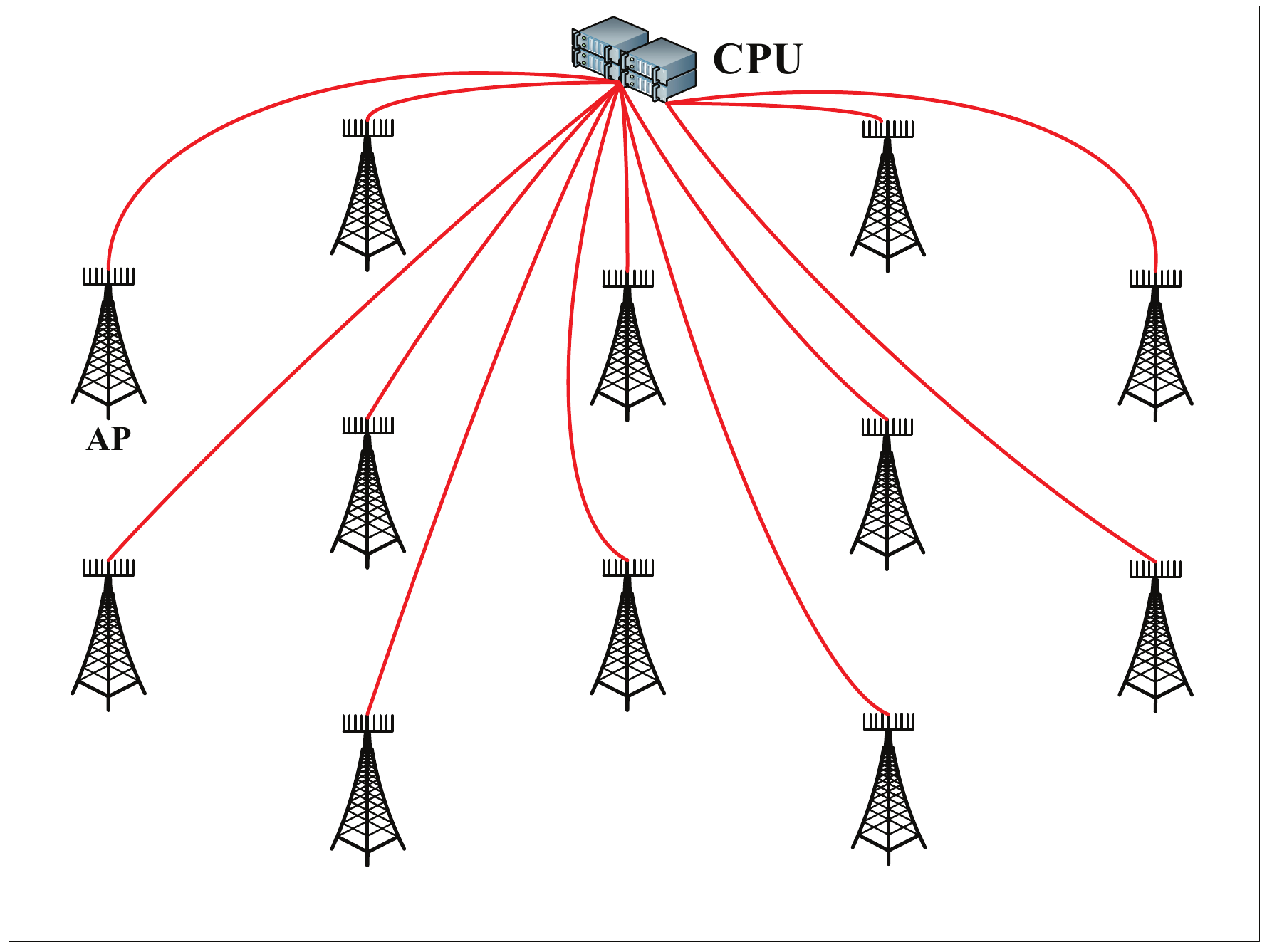}}
    \subfigure[]{\includegraphics[width=0.7\columnwidth]{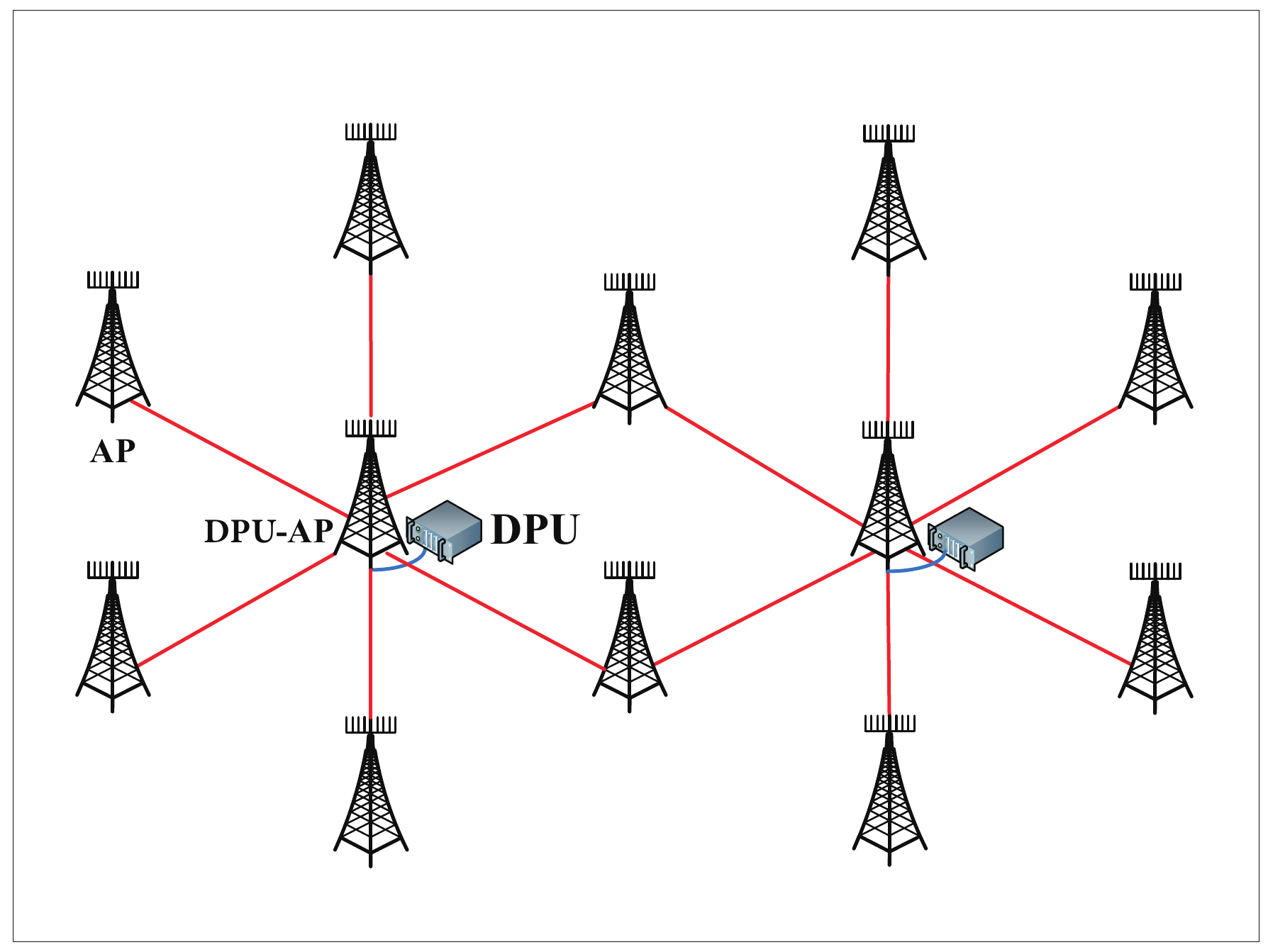}}
    \caption{Two processing paradigms for cell-free massive MIMO-based IoT: (a) Cloud computing; (b) Edge computing.}
    \label{Fig1}
    \vspace{-1mm}
\end{figure*}

\begin{itemize}

\item{\textbf{Massive access in cell-free massive MIMO-based IoT:}
We propose to employ the promising cell-free massive MIMO to support massive connectivity service in future IoT applications, and the related massive access problem is investigated.
Different from the well studied single-cell massive access~\cite{{Schepker_ISWCS'13}, {Liu_TSP'18}, {Park_SPAWC'17}, {Shao_TSP'20},{Chen_TSP'18},{Ahn_Tcom'19},{Ke_TSP'20}}, we consider a more general mMTC scenario, where UEs are distributed in a large area and multiple APs cooperate to offer a wide coverage range.
Furthermore, compared to the traditional network architecture~\cite{Chen_TWC'19}, which extends the aforementioned problem to the multi-cell massive access, the cell-free massive MIMO shows its superiority in combating inter-cell interference, the better massive access performance, and more flexible AP cooperation.}

\item{\textbf{A frame structure design for low-latency massive access:}
In grant-free massive access, for a specific frame of the uplink signals, the time-frequency radio resource is divided into multiple resource elements to transmit pilots and payload data.
We propose a frame structure tailored for massive access with OFDM transmission, where an advanced resource division strategy is considered.
Compared to the conventional frame structure in~\cite{Ke_TSP'20}, the proposed frame structure reaps a significant access latency reduction.}

\item{\textbf{Cloud computing and edge computing processing paradigms for the proposed scheme:}
We introduce two network architectures for cell-free massive MIMO systems to support the cloud computing-based and edge computing-based signal processing, respectively.
For the proposed massive access scheme, the AUD and CE performance of edge computing can approach that of cloud computing.
Moreover, edge computing has the potential to offload the computational burden from the central processing unit (CPU) in cloud computing to multiple distributed processing units (DPUs) and reduce the cooperation cost (e.g., backhaul cost and response time), but increases the price that part of APs should employ DPUs.}

\item{\textbf{SS-GAMP algorithm:}
Existing CS-based massive access schemes~\cite{Chen_TWC'19}, \cite{{Shim_CL'12}, {Abebe_CL'16}, {Wang_CL'16}, {Vaswani_TSP'16}, {Wang_CL'16_2}, {Du_JSAC'17}, {Wei_CL'17}, {Schepker_ISWCS'13}, {Liu_TSP'18}, {Park_SPAWC'17}, {Shao_TSP'20}, {Chen_TSP'18}, {Ahn_Tcom'19}, {Ke_TSP'20}} only consider the ideal processed signals with infinite-resolution quantization.
By contrast, the proposed SS-GAMP algorithm provides a general framework to achieve joint AUD and CE, where the quantization of the processed signals is considered.
Hence, for processed signals after low-resolution quantization due to the limited capacity of the wireless backhaul, the proposed algorithm has a better massive access performance than conventional algorithms in~\cite{Chen_TWC'19}, \cite{{Ke_TSP'20}}.
Moreover, we propose a weighted message refining strategy to leverage the sparsity properties of the channel matrix, which can further improve the performance in contrast to the strategy in~\cite{Ke_TSP'20}.}

\item{\textbf{SIC-based AUD and CE algorithm:}
This algorithm consists of three modules: spatial-domain AUD, angular-domain CE, and the identified UE cancellation.
These three modules are executed alternately in an iterative manner.
In contrast to the spatial-domain joint AUD and CE solutions without SIC~\cite{{Schepker_ISWCS'13}, {Liu_TSP'18}, {Park_SPAWC'17}, {Shao_TSP'20}, {Chen_TSP'18}, {Ahn_Tcom'19}}, this algorithm can dramatically reduce the access latency by further leveraging the angular domain sparsity of massive MIMO channels and the idea of SIC.}

\end{itemize}

\textit{Notations}: We use normal-face letters to denote scalars, lowercase (uppercase) boldface letters to denote column vectors (matrices).
The $(k,m)$-th element, the $k$-th row vector, and the $m$-th column vector of the matrix ${\bf H} \in {\mathbb C}^{K \times M}$ are denoted as $[{\bf H}]_{k,m}$, $[{\bf H}]_{k,:}$, and $[{\bf H}]_{:,m}$, respectively.
$\{{\bf H}_n\}_{n=1}^N$ denotes a matrix set with the cardinality of $N$ and ${\bf 0}_{K \times M}$ is the zero matrix of size ${K \times M}$.
The superscripts $(\cdot)^{\rm T}$, $(\cdot)^*$, and $(\cdot)^{\rm H}$ represent the transpose, complex conjugate, and conjugate transpose operators, respectively.
$[K]$ denotes the set of integers $\{1, 2, \cdots, K\}$, $|{\cal A}|_c$ is the cardinal number of set $\cal A$, $\emptyset$ is the empty set, and ${\rm supp}\{\cdot\}$ denotes the support set of a sparse vector or matrix.
$\lceil b \rceil$ rounds $b$ to the nearest integer greater than or equal to $b$.
${\cal U}(x;a,b)$ denotes the variable $x$ follows the uniform distribution between $a$ and $b$.
Finally, ${\cal CN}\left(x; \mu, v\right)$ denotes the complex Gaussian distribution of a random variable $x$ with mean $\mu$ and variance $v$, and ${\mathbb E}[\cdot]$ denotes statistical expectation operator.

\section{System Model}
\label{Sec:II}

In this section, we first introduce two processing paradigms for cell-free massive MIMO-based IoT.
Subsequently, we detail the procedure, the proposed frame structure, and the related signal model for massive access in cell-free massive MIMO systems.
Finally, the sparsity properties of the massive access channel matrix represented in the spatial and angular domains are illustrated.

\subsection{Proposed Cell-Free Massive MIMO-Based IoT}
\label{sec:II-A}

Consider a typical cell-free massive MIMO system to serve massive IoT UEs, where quantities of APs equipped with massive antennas cooperate in the network to serve a vast area, as illustrated in Fig.~\ref{Fig1}.
The APs are connected to the processing unit (i.e., CPU or DPU) via backhaul links, thus the received signals and information obtained at multiple APs can be jointly processed at the processing units to realize AP cooperation.
In this context, the concepts of cell and cell boundary do not exist.
Here, we consider two different processing paradigms to enable the AP cooperation for massive access: \emph{(1) Cloud computing paradigm} for centralized cooperation, where all APs will collect the signals from UEs and then transfer them to the CPU far away from the UEs, see Fig.~\ref{Fig1}(a).
The CPU will perform high computational complexity signal processing for the whole network.
Since the APs are only designed for receiving and transmitting signals, this architecture can significantly reduce the APs' cost for their large-scale deployment.
\emph{(2) Edge computing paradigm} for distributed cooperation, which offloads the signal processing from one CPU to multiple DPUs (also mobile edge computing [MEC] severs) deployed at part of the APs, and these APs are referred to as the DPU-APs, MEC-APs, or fog-APs, see Fig.~\ref{Fig1}(b).
Furthermore, other APs are connected to several adjacent DPU-APs for distributed signal processing.
In this case, the processing work is offloaded from the CPU to multiple DPUs at the corresponding DPU-APs.
Compared to the cloud computing, this paradigm can alleviate the burden on backhaul links and CPU, and support more flexible signal processing implementation.
These advantages make the edge computing-based massive access has a faster access response, while at the cost that the DPU-APs should employ extra DPUs (MEC severs).
\begin{remark}
Note that most existing cell-free massive MIMO papers consider the cell-free architecture with distributed massive MIMO configuration, i.e., each AP is equipped with one antenna or few antennas~\cite{Ngo_TWC'17}.
By contrast, this paper considers the co-located massive MIMO configuration, where each AP is equipped with massive antennas, as shown in Fig.~\ref{Fig1}.
Compared with the former one, we believe the cell-free architecture using co-located massive MIMO configuration is more practical, since most commercialized massive MIMO systems are co-located and can be easily upgraded to the cell-free architecture.
\end{remark}

\subsection{Massive Access in Cell-Free Massive MIMO Systems}
\label{sec:II-B}

The procedure of the proposed grant-free massive access scheme for cell-free massive MIMO-based IoT can be summarized as follows.

\begin{itemize}

\item{\textbf{Step 1:} During the uplink transmission phase, all active UEs directly transmit their non-orthogonal access pilot sequences and the following payload data to the APs without waiting for the access permission.}

\item{\textbf{Step 2:} Each AP collects the received signals over multiple successive time slots, and sends the collected signals to the processing unit, i.e., CPU in cloud computing or the DPUs equipped at the adjacent DPU-APs in edge computing, via backhaul links.}

\item{\textbf{Step 3:} By jointly processing the received signals from multiple APs, the processing unit performs AUD and CE for the cell-free massive MIMO-based IoT, and then the obtained AUS and corresponding channel estimates are used for subsequent data detection.}

\end{itemize}
Next, we will detail the proposed technical components.

\subsubsection{The Proposed Frame Structure Design}

\begin{figure}[t]
    \captionsetup{font={footnotesize}, name = {Fig.}, labelsep = period}
    \centering
    \includegraphics[width=0.8\columnwidth,keepaspectratio]
    {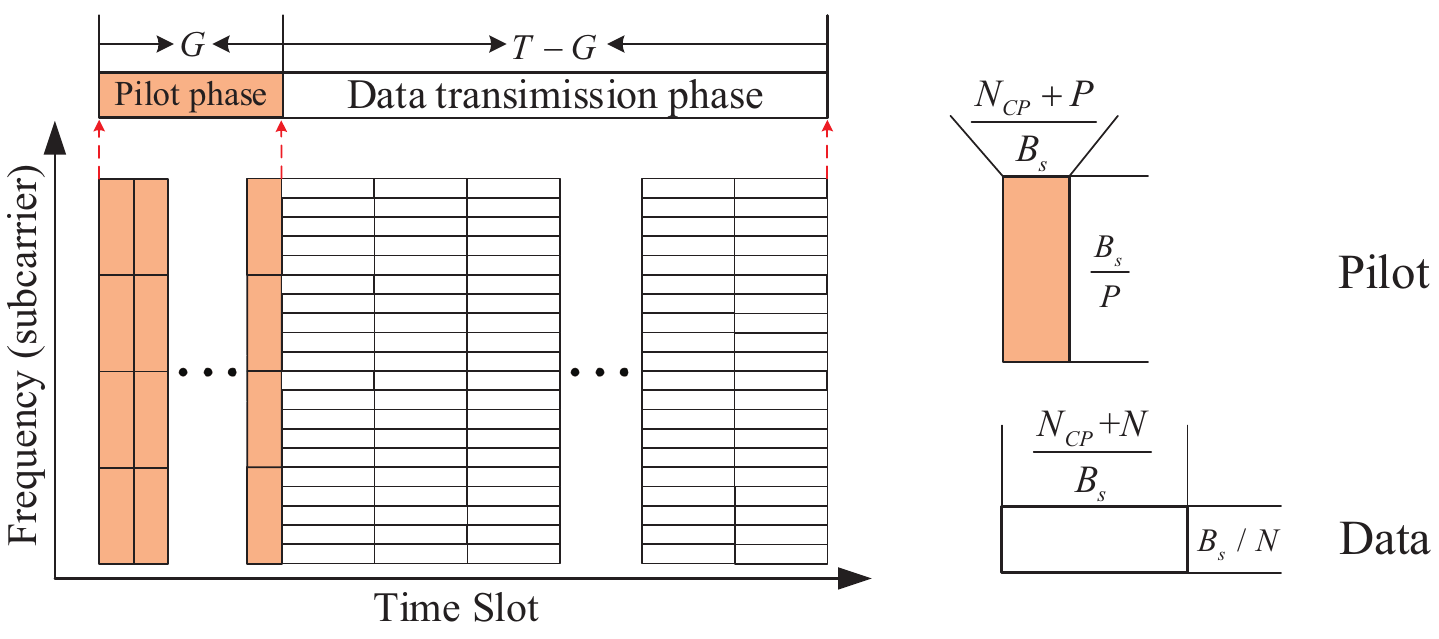}
	\caption{The proposed frame structure for the uplink transmission in grant-free massive access.}
    \label{Fig2}
\end{figure}

At the UEs side, we propose an advanced frame structure design to transmit the uplink access pilot sequence and payload data.
The proposed frame structure is illustrated in Fig.~\ref{Fig2}, where the cyclic prefix (CP)-OFDM is employed to combat time dispersive channels and the length of CP is denoted by $N_{CP}$.
By adopting OFDM, the time-frequency radio resource can be divided into multiple resource elements to convey the pilot signals and payload data.
Specifically, a frame comprising $T$ time slots is divided into two phases in the time domain, where the first $G$ time slots (i.e., pilot phase) are used to transmit access pilot signals, and the remaining $(T - G)$ time slots (i.e., data phase) are reserved for payload data transmission.
In the pilot phase, we consider the OFDM's discrete Fourier transform (DFT) length is $P = N_{CP}$, so that the subcarrier spacing is $B_s/P$ and thus each CP-OFDM symbol's duration is $(N_{CP}+P)/B_s$, where $B_s$ is the two-sided bandwidth.
In the data phase, we consider the OFDM symbol's DFT length is $N \gg P$ and thus each CP-OFDM symbol's duration is $(N_{CP}+N)/B_s$.
In grant-free massive access, the pilot signals will be used for both AUD and CE, and thus the pilot transmission latency in the proposed scheme is $G(N_{CP}+P)/B_s$.
While for the frame structure adopted by existing broadband massive access scheme in~\cite{Ke_TSP'20}, the OFDM symbol's DFT lengths in both pilot and data phases are $N$, thus the corresponding latency required is $G(N_{CP}+N)/B_s$.
Compared to the traditional frame structure, the proposed frame structure will significantly reduce the access latency as usually $P \ll N$.
For example, we consider $P = N_{CP} =64$ and $N = 2048$ in the simulations, the proposed frame structure can reap a reduction of approximately 94$\%$ in access latency.

\subsubsection{Received Signal Model at APs}

\begin{figure}[t]
    \captionsetup{font={footnotesize}, name = {Fig.}, labelsep = period}
    \centering
    \includegraphics[width=0.7\columnwidth,keepaspectratio]
    {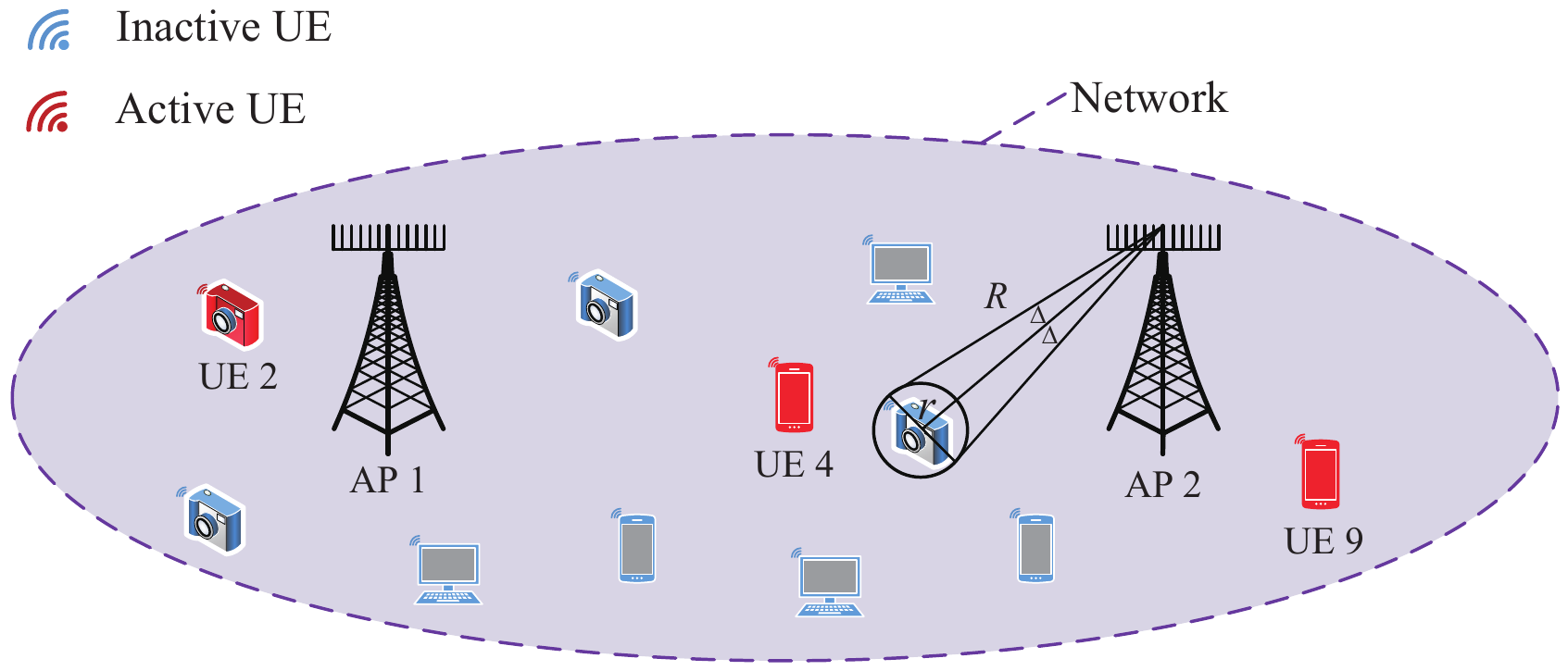}
	\caption{Sparse UE activity in massive access scenarios. A classical one-ring channel model is considered for the channels between the UEs and the massive MIMO APs.}
    \label{Fig3}
\end{figure}

We investigate a massive access problem in cell-free massive MIMO systems, where $B$ APs are employed to serve $K$ UEs, and the UEs are distributed in a vast area.
Here, $K$ is usually large (e.g., $K = 10^3$ in~\cite{Liu_TSP'18}).
Each AP is equipped with an $M_c$-antenna uniform linear array (ULA), and each UE has only one antenna without loss of generality.
Here we focus on the pilot phase with OFDM's size being $P$.
For the subchannel of the $p$-th pilot subcarrier ($1 \le p \le P$), the signal ${\bf y}_{p,b,k}^t \in \mathbb{C}^{M_c \times 1}$ received at the $b$-th AP from the $k$-th UE in the $t$-th time slot (i.e., the $t$-th OFDM symbol) is expressed as
\begin{equation}
{\bf y}_{p,b,k}^t = \sqrt{P_k}{\bf h}_{p,b,k}s_{p,k}^t + {\bf n}_{p,b}^t,
\end{equation}
where $P_k$ denotes the transmit power of the $k$-th UE, ${\bf h}_{p,b,k} \in \mathbb{C}^{M_c \times 1}$ is the subchannel associated with the $k$-th UE and the $b$-th AP, $s_{p,k}^t$ is the uplink access pilot, and ${\bf n}_{p,b}^t$ denotes the additive white Gaussian noise (AWGN).
Due to the sporadic traffic of UEs, within a given time duration, only a small number of UEs are activated and try to transmit uplink signals to the APs, as illustrated in Fig.~\ref{Fig3}.
We define an activity indicator $\alpha_k$ to indicate the UEs' activity, which equals~1 when the $k$-th UE is active and 0 otherwise.
Meanwhile, the set of active UEs is defined as ${\cal A} = \{k|\alpha_k = 1, 1 \le k \le K\}$, and the number of active UEs is denoted by $K_a = |{\cal A}|_c$.
Hence, for the $p$-th pilot subcarrier and the $t$-th time slot, the signal received at the $b$-th AP from all active UEs is given as follows
\begin{equation}\label{Eq:r_b}
{\bf y}_{p,b}^t = \sum\nolimits_{k = 1}^K \sqrt{P_k}\alpha_k{\bf h}_{p,b,k}s_{p,k}^t + {\bf n}_{p,b}^t.
\end{equation}
The channel ${\bf h}_{p,b,k}$ can be modeled as ${\bf h}_{p,b,k} = \rho_{b,k}{\widetilde {\bf h}}_{p,b,k}$, where both the large-scale fading and small-scale fading are taken into account.
Here, $\rho_{b,k}$ is the large-scale fading coefficient caused by path loss, and ${\widetilde {\bf h}}_{p,b,k}$ is the small-scale fading vector.
For the $p$-th pilot subcarrier, the subchannel between the $k$-th UE and the $b$-th AP is modeled as follows~\cite{{3GPP'15}, {Zhou_Book'07}}
\begin{equation}\label{Eq:Ch_Model}
{\widetilde {\bf h}}_{p,b,k} = \sum\nolimits_{l = 1}^{L_{b,k}} \beta_{b,k}^l{\bf a}_R\left(\phi_{b,k}^l\right)e^{-j2\pi\tau_{b,k}^lf_p},
\end{equation}
where $f_p = - \frac{B_s}2 + \frac{B_sp}P$, $L_{b,k}$ denotes the number of multi-path components (MPCs) between the $k$-th UE and the $b$-th AP, $\beta_{b,k}^l$ and $\tau_{b,k}^l$ are the complex path gain and the path delay of the $l$-th MPC, respectively.
The array response vector ${\bf a}_R\left(\phi_{b,k}^l\right)$ is given by ${\bf a}_R\left(\phi_{b,k}^l\right) = \left[1, e^{-j2\pi\phi_{b,k}^l}, \cdots, e^{-j2\pi(M_c-1)\phi_{b,k}^l}\right]^{\rm T}$, where $\phi_{b,k}^l = \frac{\widetilde d}\lambda{\rm sin}\left(\varphi _{b,k}^l\right)$.
Here, $\varphi _{b,k}^l$ is the angle of arrival (AOA) observed at the AP side, $\lambda$ denotes the wavelength, and the antenna spacing ${\widetilde d} = \lambda/2$ is considered.

\subsection{Sparsity Properties of the Massive Access Channel Matrix}
\label{Sec:II-C}

Define ${\bf H}_{p,b} = \left[\sqrt{P_1}\alpha_1{\bf h}_{p,b,1}, \cdots, \sqrt{P_K}\alpha_K{\bf h}_{p,b,K}\right]^{\rm T} \in {\mathbb C}^{K \times M_c}$ as the massive access channel matrix between all UEs and the $b$-th AP at the $p$-th pilot subcarrier.
In this section, we first present the structured sparsity of the spatial-domain channel matrices $\left\{ {\bf H}_{p,b} \right\}_{p=1}^P, \forall b$.
Furthermore, by representing the MIMO channels in the virtual angular domain, the structured sparsity of the angular-domain channel matrices $\left\{ {\bf W}_{p,b} \right\}_{p=1}^P, \forall b$, is further illustrated.

\subsubsection{Spatial-Domain Structured Sparsity}

\begin{figure}[t]
    \captionsetup{font={footnotesize}, name = {Fig.}, labelsep = period}
    \centering
    \includegraphics[width=1\columnwidth,keepaspectratio]
    {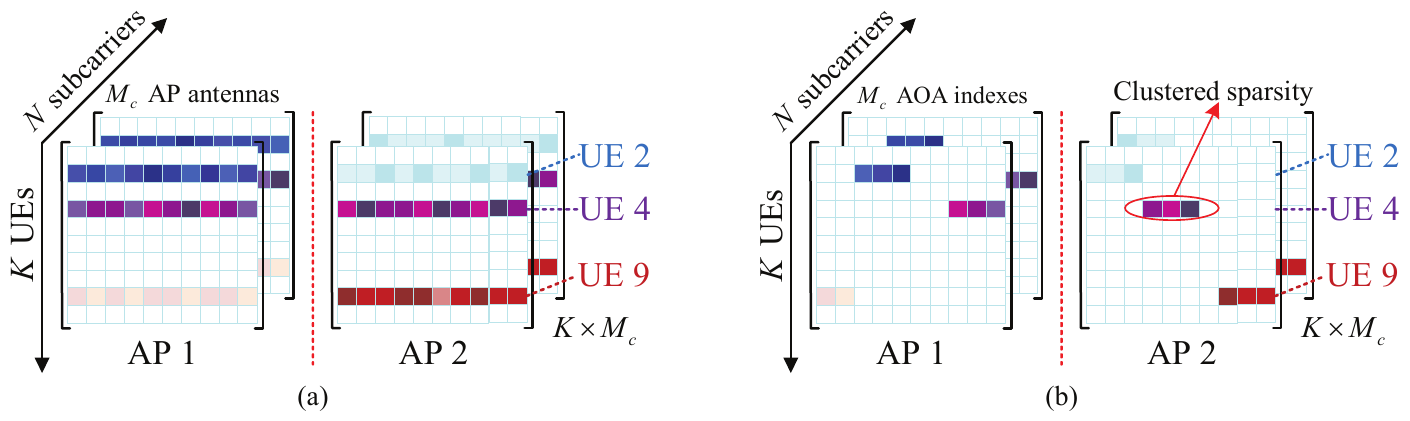}
	\caption{The massive access channel matrix exhibits two forms of structured sparsity in the spatial and angular domains: (a) Spatial-domain structured sparsity due to sparse UE activity; (b) Angular-domain structured sparsity due to the limited angular spread of the MPCs. A darker color denotes a higher channel gain.}
    \label{Fig4}
\end{figure}

For a typical massive access scenario, only a small number of UEs out of total $K$ UEs are active, i.e., most of $\alpha_k, \forall k$ are equal to 0.
Thus, the channel vector $\left[{\bf H}_{p,b}\right]_{:,m}$ observed at the $m$-th receive antenna of the $b$-th AP is sparse, i.e.,
\begin{equation}\label{Eq:Sparse_SPa_1}
\left|{\rm supp}\left\{\left[{\bf H}_{p,b}\right]_{:,m}\right\}\right|_c = K_a \ll K.
\end{equation}
Moreover, given the UE activity, i.e., the value of $\alpha_k$, all elements of the $k$-th row of $\{ {\bf H}_{p,b} \}_{p=1}^P$ will be zero or non-zero simultaneously.
Therefore, the sparsity pattern (\ref{Eq:Sparse_SPa_1}) can be simultaneously observed at different AP antennas and different subcarriers, which can be expressed as
\begin{equation}\label{Eq:Sparse_SPa_2}
{\rm supp}\{\left[{\bf H}_{p,b}\right]_{:,1}\} = {\rm supp}\{\left[{\bf H}_{p,b}\right]_{:,2}\} = \cdots = {\rm supp}\{\left[{\bf H}_{p,b}\right]_{:,M_c}\},
\end{equation}
and
\begin{equation}\label{Eq:Sparse_SPa_3}
{\rm supp}\left\{{\bf H}_{1,b}\right\} = {\rm supp}\left\{{\bf H}_{2,b}\right\} = \cdots = {\rm supp}\left\{{\bf H}_{P,b}\right\},
\end{equation}
respectively.
We refer to the structured sparsity in (\ref{Eq:Sparse_SPa_1})-(\ref{Eq:Sparse_SPa_3}) as the spatial-domain structured sparsity of $\{ {\bf H}_{p,b} \}_{p=1}^P$.
Particularly, the signals of active UEs can be received by all APs, and this structured sparsity caused by sporadic UEs' traffic would be the same for different APs.
On the other hand, due to the large-scale fading caused by path loss, the channel strength from a specific active UE to far away APs can be approximate zero.
Hence, the channel matrices between UEs and different APs, i.e., $\{ {\bf H}_{p,b} \}_{p=1}^P, \forall b$, exhibit approximate common sparsity pattern.
To illustrate this structured sparsity more explicitly, we provide an example in Fig.~\ref{Fig3} and Fig.~\ref{Fig4}(a), where we assume that $K_a = 3$ active UEs out of $K = 10$ total UEs access the network and each AP is equipped with $M_c = 10$ antennas.
Given the locations of active UEs and APs described in Fig.~\ref{Fig3}, the $4$-th row vectors in both channel matrices $\{ {\bf H}_{p,1} \}_{p=1}^P$ and $\{ {\bf H}_{p,2} \}_{p=1}^P$ (corresponding to the $4$-th UE in the active state in Fig.~\ref{Fig4}) have large gain (strong common support).
However, due to the large path loss gap, for the $2$-th (or $9$-th) UE in the active state, only the $2$-th (or $9$-th) row vectors in $\{ {\bf H}_{p,1} \}_{p=1}^P$ (or $\{ {\bf H}_{p,2} \}_{p=1}^P$) have the sufficiently large gain while those in $\{ {\bf H}_{p,2} \}_{p=1}^P$ (or $\{ {\bf H}_{p,1} \}_{p=1}^P$) can be negligible, which can be illustrated in Fig.~\ref{Fig4}(a).

\subsubsection{Angular-Domain Structured Sparsity}

By representing the massive MIMO channels in the virtual angular domain, we can find some additional sparsity properties of the massive access channel matrix.
Specifically, the angular-domain massive MIMO channel between the $k$-th UE and the $b$-th AP at the $p$-th pilot subcarrier can be represented as
\begin{equation}\label{Eq:Spa_to_Ang}
{\widetilde {\bf w}_{p,b,k}} = {{\bf A}_R^{\rm H}}{\widetilde {\bf h}_{p,b,k}},
\end{equation}
where the transformation matrix ${\bf A}_R \in \mathbb{C}^{M_c \times M_c}$ at the AP side is a unitary matrix.
Here, ${\bf A}_R$ depends on the geometry of the array, which becomes the DFT matrix for a ULA when ${\widetilde d} = \lambda /2$~\cite{Zhou_Book'07}.
For the practical implementation of the network, the APs are usually deployed at high elevation with few scatterers around, whereas the UEs are typically distributed at low elevation in a local rich scattering environment far from the APs~\cite{Gao_TSP'15}.
We model this typical scenario as the classical one-ring channel model~\cite{Nam_JSTSP'14}, as illustrated in Fig.~\ref{Fig3}.
Here, we assume a UE is located in a rich scattering environment within a radius of $r$, and the distance between the UE and AP is $R$, so the angular spread observed at the AP is given as
\begin{equation}\label{Eq:Ang_Spread}
\Delta  \approx {\rm arctan}\left(r/R\right).
\end{equation}
Hence, the sparsity level of the angular-domain channels is proportional to $\Delta$, and it is expected to be far less than $M_c$ as usually $R \gg r$.
This indicates the virtual angular-domain sparsity of massive MIMO channels, i.e.,
\begin{equation}\label{Eq:Sparse_Ang_1}
\left|{\rm supp}\left\{\widetilde {\bf w}_{p,b,k}\right\}\right|_c \ll M_c,
\end{equation}
and this sparsity is clustered, as illustrated in Fig.~\ref{Fig4}(b).
Furthermore, as the scattering environment for all subchannels within the bandwidth remains unchanged, the angular spreads of all subchannels are very similar.
Hence, all subchannels have a common sparsity pattern as
\begin{equation}\label{Eq:Sparse_Ang_2}
{\rm supp}\left\{\widetilde {\bf w}_{1,b,k}\right\} = {\rm supp}\left\{\widetilde {\bf w}_{2,b,k}\right\} =  \cdots  = {\rm supp}\left\{\widetilde {\bf w}_{P,b,k}\right\}.
\end{equation}
We refer to the structured sparsity in (\ref{Eq:Sparse_Ang_1}) and (\ref{Eq:Sparse_Ang_2}) as the angular-domain structured sparsity.
Define the virtual angular-domain channel matrix as ${\bf W}_{p,b} = {{\bf H}_{p,b}}{{\bf A}_R^*} = \left[\sqrt{P_1}\alpha_1{{\bf w}_{p,b,1}}, \cdots, \sqrt{P_K}\alpha_K{{\bf w}_{p,b,K}}\right]^{\rm T}$, where ${\bf w}_{p,b,k} = \rho_{b,k}{\widetilde {\bf w}_{p,b,k}}$.
By combining the sparse UE activity and the angular-domain structured sparsity, we further have $\left|{\rm supp}\left\{\left[{\bf W}_{p,b}\right]_{:,m}\right\}\right|_c \ll K_a$, and
\begin{equation}\label{Eq:Sparse_Ang_3}
{\rm supp}\left\{{\bf W}_{1,b}\right\} = {\rm supp}\left\{{\bf W}_{2,b}\right\} = \cdots = {\rm supp}\left\{{\bf W}_{P,b}\right\}.
\end{equation}
An illustration of the structured sparsity of $\{ {\bf W}_{p,b} \}_{p=1}^{P}, \forall b$ is also provided in Fig.~\ref{Fig4}(b).

\begin{remark}
The aforementioned angular-domain structured sparsity of massive MIMO channels is valid even for sub-6 GHz systems~\cite{Lin_CL'17}.
Note that our work considers the cell-free network based on co-located massive MIMO configuration, rather than distributed massive MIMO~\cite{Ngo_TWC'17} whose angular-domain sparsity does no exist.
\end{remark}

These two forms of structured sparsity will be leveraged to facilitate the design of AUD and CE algorithm in the remainder of this paper.
Specifically, the spatial-domain approximate common sparsity can be exploited to enhance the AUD performance, while the angular-domain enhanced sparsity can be utilized to improve the CE performance.
Hence, we perform AUD based on the spatial-domain channel model and perform CE for the identified UEs based on the angular-domain channel model.

\section{Cloud Computing-Based and Edge Computing-Based Massive Access}
\label{Sec:III}

This section details the problem formulations at the receiver for massive access based on cloud computing and edge computing paradigms, respectively.
In this paper, we adopt a grant-free massive access protocol to avoid complicated access scheduling, where the transmit frame structure proposed in Section \ref{sec:II-B} is employed.
Here, we assume the frame length is far smaller than the channel coherence time, and the activity of the UEs during the channel coherence time remains unchanged.
In grant-free massive access, the set of active UEs and the corresponding CSI have to be acquired for the subsequent coherent data detection.
In the pilot phase, for the $b$-th AP and the $p$-th pilot subcarrier, the pilot signals received in $G$ successive time slots are collected as
\begin{equation}\label{Eq:R_b}
\begin{aligned}
{\bf Y}_{p,b} & = \sum\nolimits_{k = 1}^K {\bf s}_{p,k}\sqrt{P_k}\alpha_k{\bf h}_{p,b,k}^{\rm T} + {\bf N}_{p,b}\\
              & = {\bf S}_p{\bf H}_{p,b} + {\bf N}_{p,b},\; \forall p \in [P]\; {\rm and}\; \forall b \in [B],
\end{aligned}
\end{equation}
where ${\bf Y}_{p,b} = \left[{\bf y}_{p,b}^1, \cdots, {\bf y}_{p,b}^G\right]^{\rm T} \in {\mathbb C}^{G \times M_c}$ and the received signal ${\bf y}_{p,b}^t$ is given in (\ref{Eq:r_b}).
Furthermore, ${\bf s}_{p,k} = [s_{p,k}^1, \cdots, s_{p,k}^G]^{\rm T} \in {\mathbb C}^{G \times 1}$ is the access pilot sequence of the $k$-th UE at the $p$-th pilot subcarrier, ${\bf S}_{p} = \left[{\bf s}_{p,1}, \cdots, {\bf s}_{p,K}\right] \in {\mathbb C}^{G \times K}$ is the pilot matrix.
Here, the pilot of the $k$-th UE at the $p$-th pilot subcarrier is given as ${\cal CN} \sim \left( s_{p,k}^t ; 0,1\right)$, and the pilots at different pilot subcarriers are different for achieving diversity~\cite{Gao_TSP'15}.
Finally, ${\bf H}_{p,b},\; \forall p \in [P]$, denotes the massive access channel matrix between all UEs and the $b$-th AP, and ${\bf N}_{p,b} = \left[{\bf n}_{p,b}^1, \cdots, {\bf n}_{p,b}^G\right]^{\rm T}$.
Based on (\ref{Eq:R_b}), the AUD problem is to estimate $\alpha_k, \forall k \in [K]$, i.e., find the indices of non-zero rows of $\{ {\bf H}_{p,b} \}_{p=1}^P, \forall b \in [B]$; on the other hand, the CE problem is to estimate ${\bf h}_{p,b,k}$ for $\forall k \in {\cal A}$, i.e., the related row coefficients of $\{ {\bf H}_{p,b} \}_{p=1}^P, \forall b \in [B]$.
Therefore, these two problems can be jointly solved by estimating $\{ {\bf H}_{p,b} \}_{p=1}^P$ based on the known ${\bf S}_p$ and ${\bf Y}_{p,b}, \forall b \in [B]$.

\subsection{Cloud Computing-Based Massive Access}
\label{Sec:III-A}

For cloud computing paradigm, quantities of APs are distributed in a large area and cooperate at the CPU through backhaul links.
Here, the APs are only designed for receiving and transmitting signals, thus the corresponding AUD and CE are centrally processed at the CPU.
Considering the limited capacity of wireless backhaul links, the signals received at the APs are first quantized and then transmitted via backhaul links\footnote{Especially for the widely used wireless backhaul with limited capacity, the higher resolution of quantization benefits the better massive access performance but at the cost of larger backhaul latency.} to the CPU, i.e., $\forall p \in [P]$ and $\forall b \in [B]$,
\begin{equation}\label{Eq:Sig_quan}
{\overline {\bf Y}_{p,b}} = \psi_b\left({\bf Y}_{p,b}\right) = \psi_b\left({\bf S}_p{\bf H}_{p,b} + {\bf N}_{p,b}\right),
\end{equation}
where $\psi_b\left(\cdot\right)$ is the complex-valued quantizer at the $b$-th AP.
The quantizer is applied to the received signal element-wisely, and the real and imaginary parts are quantized separately.
Here, we consider a uniform codebook for quantization,
\begin{equation}\label{Eq:Cloud_Process_Spa}
{\cal C}_b = \left\{ -\frac{2^Q-1}2\Delta_b,\cdots,\frac{2^Q-1}2\Delta_b\right\},
\end{equation}
where $Q$ is the number of quantization bits, $\Delta_b = \left(y_b^{\rm max} - y_b^{\rm min}\right)/2^Q$, $y_b^{\rm max}$ and $y_b^{\rm min}$ are the maximum and the minimum real values of both real and imaginary parts of $\left\{{\bf Y}_{p,b}\right\}_{p=1}^P$, respectively.
At the CPU, the quantized received signals from all APs are concentrated as $\forall p \in[P]$,
\begin{equation}\label{Eq:Cloud_Process_Spa}
{\overline {\bf Y}_p} = \left[{\overline {\bf Y}_{p,1}}, {\overline {\bf Y}_{p,2}}, \cdots, {\overline {\bf Y}_{p,B}}\right]
          = {\bf S}_p{\bf H}_p + {\bf N}_p^q + {\bf N}_p,
\end{equation}
where ${\bf N}_p^q$ denotes the quantization error, ${\bf H}_p \in \mathbb{C}^{K \times M}$ is expressed as ${\bf H}_p = \left[{\bf H}_{p,1}, {\bf H}_{p,2}, \cdots, {\bf H}_{p,B}\right]$, $M = BM_c$, and ${\bf N}_p = \left[{\bf N}_{p,1}, {\bf N}_{p,2}, \cdots, {\bf N}_{p,B}\right]$.
In stark contrast to the standard linear model (SLM) with infinite-resolution quantization widely used in~\cite{{Ngo_TWC'17},{He_TWC'18}}, the model (\ref{Eq:Cloud_Process_Spa}) is a generalized linear model (GLM) due to the nonlinear measurements.
By exploiting the sparse UE activity, the AUD problem based on (\ref{Eq:Cloud_Process_Spa}) is formulated as a GLM-based CS problem, where we seek to recover the sparse channel matrices $\{{\bf H}_p\}_{p=1}^P$ from the quantized measurements $\left\{{\overline {\bf Y}_p}\right\}_{p=1}^P$.
Meanwhile, the spatial-domain structured sparsity of $\{{\bf H}_p\}_{p=1}^P$, as described in Section \ref{Sec:II-C} and illustrated in Fig. \ref{Fig4}(a), can be exploited to improve the detection performance.

On the other hand, by representing the massive MIMO channels in the virtual angular domain, we can further transform (\ref{Eq:Sig_quan}) into
\begin{equation}\label{Eq:Rb_Ang_1}
{\bf R}_{p,b} = {\overline {\bf Y}_{p,b}}{\bf A}_R^* = {\bf S}_p{\bf W}_{p,b} + {\overline {\bf N}_{p,b}^q} + {\overline {\bf N}_{p,b}},
\end{equation}
where ${\overline {\bf N}}_{p,b}^q = {\bf N}_{p,b}^q{\bf A}_R^*$ and ${\overline {\bf N}}_{p,b} = {\bf N}_{p,b}{\bf A}_R^*$.
Thus, the (\ref{Eq:Cloud_Process_Spa}) at the CPU can be also expressed as
\begin{equation}\label{Eq:Cloud_Process_Ang}
{\bf R}_p = \left[{\bf R}_{p,1}, {\bf R}_{p,2}, \cdots, {\bf R}_{p,B}\right]
          = {\bf S}_p{\bf W}_p + {\overline {\bf N}_p^q} + {\overline {\bf N}}_p,
\end{equation}
where ${\bf W}_p = \left[{\bf W}_{p,1}, {\bf W}_{p,2}, \cdots, {\bf W}_{p,B}\right]$ and ${\overline {\bf N}}_p = \left[{\overline {\bf N}}_{p,1}, {\overline {\bf N}}_{p,2}, \cdots, {\overline {\bf N}}_{p,B}\right]$.
With the estimate of AUS based on (\ref{Eq:Cloud_Process_Spa}), denoted as ${\widehat {\cal A}}$, the CE problem based on (\ref{Eq:Cloud_Process_Ang}) is  equivalent to solving the following CS problem
\begin{equation}\label{Eq:Rb_Ang_2}
{\bf R}_{p} = \left[{\bf S}_p\right]_{:,{\widehat {\cal A}}} \left[{\bf W}_{p}\right]_{{\widehat {\cal A}},:} + {\widetilde {\bf N}_{p}},
\end{equation}
where ${\widetilde {\bf N}}_p$ includes the aggregated AWGN, quantization error, and estimation error of AUD.
By leveraging the angular-domain structured sparsity of $\{{\bf W}_p\}_{p=1}^P$, as described in Section \ref{Sec:II-C} and illustrated in Fig. \ref{Fig4}(b), the CSI estimates of the UEs identified in (\ref{Eq:Cloud_Process_Spa}) can be further refined.

Hence, by leveraging the two forms of structured sparsity the channel matrix, the AUD and CE problems based on cloud computing paradigm are equivalent to solving the CS problems in (\ref{Eq:Cloud_Process_Spa}) and (\ref{Eq:Rb_Ang_2}), respectively, i.e., detecting the non-zero rows of $\{{\bf H}_p\}_{p=1}^P$ and estimating the corresponding row coefficients of $\{{\bf W}_p\}_{p=1}^P$.

\subsection{Edge Computing-Based Massive Access}
\label{Sec:III-B}

For edge computing paradigm, the central processing at the CPU is offloaded to the edge of the network, as illustrated in Fig. \ref{Fig1}(b).
Specifically, a part of the APs, termed as DPU-APs, are equipped with the DPUs or MEC severs having the storage and the computing capabilities.
Hence, the signals received at multiple APs are jointly processed at the adjacent DPU-APs.
We will further explain this distributed processing strategy from a DPU-AP centric perspective.
Specifically, for a specific DPU-AP, its DPU will collect the signals received locally and from the $(N_{co}-1)$ nearest APs for the distributed processing.
Here, $N_{co}$ is the number of APs for cooperation, which includes one DPU-AP and $(N_{co}-1)$ conventional APs without DPU.
Assume there are $I$ DPU-APs in the network, the signals received at the $i$-th DPU-AP are organized as
\begin{equation}\label{Eq:Edge_Spa}
\begin{aligned}
{\overline {\bf Y}_{p,i}} &= \left[{\overline {\bf Y}_{p,{\cal B}_i(1)}}, {\overline {\bf Y}_{p,{\cal B}_i(2)}}, \cdots, {\overline {\bf Y}_{p,{\cal B}_i(N_{co}})}\right]\\
              &= {\bf S}_p\left[{\bf H}_p\right]_{:,{\cal M}_i} + \left[{\bf N}_p^q\right]_{:,{\cal M}_i} + \left[{\bf N}_p\right]_{:,{\cal M}_i},
\end{aligned}
\end{equation}
where ${\cal B}_i$ denotes the set of APs cooperate on the $i$-th DPU-AP, ${\cal B}_i(n)$ is the $n$-th element of ${\cal B}_i$, and the column index set ${\cal M}_i$ is defined as ${\cal M}_i = \left\{m|m=(b-1)M_c+1:bM_c, \forall b \in {\cal B}_i\right\}$.
Meanwhile, the spatial-domain channel model (\ref{Eq:Edge_Spa}) can be further represented in the angular domain as
\begin{equation}\label{Eq:Edge_Ang}
\begin{aligned}
{\bf R}_{p,i} &= \left[{\bf R}_{p,{\cal B}_i(1)}, {\bf R}_{p,{\cal B}_i(2)}, \cdots, {\bf R}_{p,{\cal B}_i(N_{co})}\right]\\
              &= {\bf S}_p\left[{\bf W}_p\right]_{:,{\cal M}_i} + \left[{\overline {\bf N}}_p^q\right]_{:,{\cal M}_i} +\left[{\overline {\bf N}}_p\right]_{:,{\cal M}_i}.
\end{aligned}
\end{equation}
For all DPU-APs, i.e., $\forall i \in [I]$, by exploiting the sparsity properties of the channel matrix, the AUD and CE problems based on edge computing paradigm are equivalent to solving the CS problems in (\ref{Eq:Edge_Spa}) and (\ref{Eq:Edge_Ang}), respectively.

\section{Proposed Active User Detection and Channel Estimation Algorithm}
\label{Sec:IV}

As described in Section \ref{Sec:III}, the AUD and CE problems for grant-free massive access are formulated as the CS problems.
In this section, we first develop a SS-GAMP algorithm to realize the related sparse signal recovery with quantized measurements.
On this basis, a SIC-based AUD and CE algorithm is further proposed.
Here, the explanations of the proposed algorithms are based on the cloud computing as an example, which can be easily extended to the edge computing.

\subsection{SS-GAMP Algorithm}
\label{Sec:IV-A}

For the CS problem with quantized measurements, we adopt the unified Bayesian inference framework proposed in~\cite{Meng_SPL'18}, which can iteratively reduce the GLM problem to a series of SLM problems.
Moreover, based on the message passing theory and employing the low-complexity heuristics for approximating the messages, we develop a SS-GAMP algorithm to reap both the better performance than greedy methods~\cite{{Gao_TSP'15}} and the lower complexity than conventional message passing algorithms~\cite{Kschi_TIT'01}.
To simplify the derivations, we focus on the spatial-domain channel model (\ref{Eq:Cloud_Process_Spa}) and the $p$-th pilot subcarrier first.
The acquired key steps of the proposed algorithm can be easily extended to the angular-domain channel model (\ref{Eq:Cloud_Process_Ang}) and multiple pilot subcarriers cases.
Furthermore, for notational simplicity, the index $p$ in ${\overline {\bf Y}_p}$, ${\bf S}_p$, and ${\bf H}_p$ is dropped and will be reused when the multiple pilot subcarriers case is considered.

\begin{figure}[t]
    \captionsetup{font={footnotesize, color = black}, name = {Fig.}, labelsep = period}
    \centering
    \includegraphics[width=0.8\columnwidth,keepaspectratio]
    {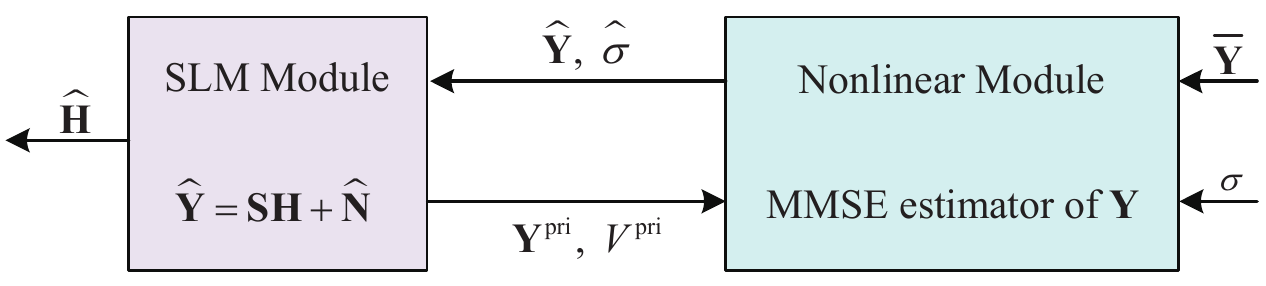}
    \caption{Block diagram of the proposed SS-GAMP algorithm.}
    \label{Fig5}
\end{figure}

The block diagram of the proposed SS-GAMP algorithm is illustrated in Fig. \ref{Fig5}, which comprises two modules: nonlinear module and SLM module.
Based on the quantized received signal ${\overline {\bf Y}}$ and the noise variance $\sigma$, nonlinear module performs minimum mean square error (MMSE) estimate of the linear received signal ${\bf Y} = {\bf S}{\bf H} + {\bf N}$, and the corresponding posterior mean and variance are denoted by ${\bf Y}^{\rm post}$ and $V^{\rm post}$, respectively.
The extrinsic messages of nonlinear module, i.e., the equivalent linear measurement ${\widehat {\bf Y}}$ and noise variance ${\widehat \sigma}$, form the input of SLM module.
In SLM module, the concerned GLM problem has been transformed into an equivalent SLM problem as
\begin{equation}\label{Eq:SLM}
{\widehat {\bf Y}} = {\bf S}{\bf H} + {\widehat {\bf N}},
\end{equation}
where the variance of the equivalent noise ${\widehat {\bf N}}$ is given as ${\widehat \sigma}$.
Hence, SLM module employs the SLM-based AMP algorithm to estimate the channel matrix ${\bf H}$, and its extrinsic messages, i.e., ${\bf Y}^{\rm pri}$ and $V^{\rm pri}$, are passed to nonlinear module as the a priori information of ${\bf Y}$.
These two modules are executed alternately in a turbo manner until convergence.

\subsubsection{Nonlinear Module}

The posterior probability of ${\bf Y}$ is expressed as
\begin{equation}
p\left({\bf Y}|{\overline {\bf Y}}\right) \propto p\left({\overline {\bf Y}}|{\bf Y}\right){\cal CN}\left({\bf Y};{\bf Y}^{\rm pri},V^{\rm pri}\right),
\end{equation}
where $p\left({\overline {\bf Y}}|{\bf Y}\right)$ is the likelihood function.
Since the processed signals are quantized element-wisely, we can compute $p\left({\bf Y}|{\overline {\bf Y}}\right)$ element-wisely, and the real and imaginary parts are calculated separately.
Furthermore, as the quantization codebooks of different APs are different, the signals from different APs are also processed separately.
According to the derivations in~\cite{Wen_TSP'16}, the posterior mean and variance of the real part of ${\bf Y}_{p,b}$ are finally given as
\begin{equation}\label{Eq:quan_post_mean}
y_{b,g,m}^{\rm post} = y_{b,g,m}^{\rm pri} + \frac{{\rm sign}\left({\overline y_{b,g,m}}\right){V^{\rm pri}}}{\sqrt{2\left(\sigma + {V^{\rm pri}}\right)}}\left( \frac{\phi\left(\eta_1\right)-\phi\left(\eta_2\right)}{\Phi\left(\eta_1\right)-\Phi\left(\eta_2\right)} \right),
\end{equation}
\begin{equation}\label{Eq:real_var}
\begin{aligned}
V^{\rm post} &= \frac{V^{\rm pri}}{2} - \frac{(V^{\rm pri})^2}{2\left(\sigma+V^{\rm pri}\right)}\\
             &\times \left(\frac{\eta_1\phi\left(\eta_1\right) - \eta_2\phi\left(\eta_2\right)} {\Phi\left(\eta_1\right)-\Phi\left(\eta_2\right)}  + \left(\frac{\phi\left(\eta_1\right)-\phi\left(\eta_2\right)}{\Phi\left(\eta_1\right)-\Phi\left(\eta_2\right)}\right)^2\right),
\end{aligned}
\end{equation}
where $\phi(\cdot)$ and $\Phi(\cdot)$ are the cumulative distribution function and the probability density function of the standard normal distribution, respectively.
In (\ref{Eq:quan_post_mean}) and (\ref{Eq:real_var}), $\eta_1$ and $\eta_2$ are defined as
\begin{equation}\label{Eq:eta_1}
\eta_1 = \frac{{\rm sign}\left({\overline y_{b,g,m}}\right)\! -\! {\rm min}\left\{|{\overline y_{b,g,m}}\!-\!\Delta_b/2|,|{\overline y_{b,g,m}}\!+\!\Delta_b/2|\right\}}{\sqrt{\frac{\sigma+V^{\rm pri}}{2}}},
\end{equation}
\begin{equation}\label{Eq:eta_2}
\eta_2 = \frac{{\rm sign}\left({\overline y_{b,g,m}}\right)\! -\! {\rm max}\left\{|{\overline y_{b,g,m}}\!-\!\Delta_b/2|,|{\overline y_{b,g,m}}\!+\!\Delta_b/2|\right\}}{\sqrt{\frac{\sigma+V^{\rm pri}}{2}}},
\end{equation}
respectively. For ease of notation, we have abused $y_{b,g,m}^{\rm post}$, $y_{b,g,m}^{\rm pri}$, and ${\overline y_{b,g,m}}$ to denote the real part of these variables, and the imaginary part can be computed analogously.
Furthermore, the extrinsic messages of nonlinear module are computed as
\begin{align}
{\widehat \sigma} &= \frac{V^{\rm post}V^{\rm pri}} {V^{\rm pri} - V^{\rm post}},\label{Eq:ext_mean1}\\
{\widehat {\bf Y}} &= {\widehat \sigma}\left({\bf Y}^{\rm post}/{V^{\rm post}} - {\bf Y}^{\rm pri}/{V^{\rm pri}}\right).\label{Eq:ext_mean_nonlin}
\end{align}
Note that ${\widehat {\bf Y}}$ and ${\widehat \sigma}$ are actually the equivalent measurement of $\left({\bf S}{\bf H} + {\bf N}\right)$ and the equivalent noise variance, respectively.

\subsubsection{SLM Module}

In this module, the quantized CS problem (\ref{Eq:Cloud_Process_Spa}) has been transformed into the conventional SLM problem, as in (\ref{Eq:SLM}).
Thus, the AMP algorithm proposed in~\cite{Ke_TSP'20}, which are designed for CS problems with linear measurements, can be directly applied to acquire the estimate of ${\bf H}$.
Due to the limited paper length, here we only clarify the key steps, and please refer to~\cite{Ke_TSP'20} for more details.
Based on the derivations in~\cite{Ke_TSP'20}, the AMP algorithm can be explained intuitively.
In the large system limit, i.e., as $K \rightarrow \infty$, while $\gamma = K_a/K$ and $\kappa = G/K$ are fixed, the AMP algorithm decouples the matrix estimation problem based on (\ref{Eq:SLM}) into $KM$ scalar estimation problems, as $\forall k \in [K]$ and $\forall m \in [M]$,
\begin{equation}\label{Eq:AMP_Decouple}
{\widehat {\bf Y}} = {\bf S}{\bf H} + {\widehat {\bf N}} \to A_{k,m}^q = h_{k,m} + {\widehat n_{k,m}^q},
\end{equation}
where $A_{k,m}^q \sim {\cal CN}\left( A_{k,m}^q; h_{k,m}, B_{k,m}^q \right)$ is the equivalent measurement of $h_{k,m}$ obtained in the $q$-th iteration of AMP algorithm, and ${\widehat n_{k,m}^q} \sim {\cal CN}\left( {\widehat n_{k,m}^q}; 0, B_{k,m}^q \right)$ denotes the effective noise.
The effective noise includes AWGN and the estimation error of $h_{k,m}$ in the $q$-th iteration.
In this way, the posterior distribution of $h_{k,m}$, $\forall k,m$, can be approximated as
\begin{equation}\label{Eq:Post_Approx1}
\begin{aligned}
p\left(h_{k,m}|{\widehat {\bf Y}}\right) &\approx p\left(h_{k,m}|A_{k,m}^q, B_{k,m}^q\right)\\
                                         &\approx \frac{1}{F_1}p_0\left(h_{k,m}\right){\cal CN}\left(h_{k,m}; A_{k,m}^q, B_{k,m}^q\right),
\end{aligned}
\end{equation}
where ${F_1}$ is a normalization factor and $p_0\left(h_{k,m}\right)$ denotes the a priori distribution of $h_{k,m}$.
In (\ref{Eq:Post_Approx1}), $B_{k,m}^q$ and $A_{k,m}^q$ are updated as follows
\begin{align}
B_{k,m}^q &= \left[\sum\nolimits_{g=1}^G \frac{\left|s_{g,k}\right|^2}{{\widehat \sigma} + C_{g,m}^q}\right]^{-1}, \label{Eq:Var_Update1} \\
A_{k,m}^q &= {\hat h_{k,m}^q} + B_{k,m}^q\sum\nolimits_{g=1}^G{\frac{s_{g,k}^*\left({\widehat y_{g,m}} - D_{g,m}^q\right)}{{\widehat \sigma} + C_{g,m}^q}}, \label{Eq:Var_Update2}
\end{align}
and $C_{g,m}^q $ and $D_{g,m}^q$ are updated as follows
\begin{align}
C_{g,m}^q &= \sum\nolimits_{k=1}^K{\left|s_{g,k}\right|^2{v_{k,m}^q}}, \label{Eq:Fac_Update1}\\
D_{g,m}^q &= \sum\nolimits_{k=1}^K{s_{g,k}{\hat h_{k,m}^q} - \frac{C_{g,m}^q}{{\widehat \sigma} + C_{g,m}^{q - 1}}\left({\widehat y_{g,m}} - D_{g,m}^{q - 1}\right)}, \label{Eq:Fac_Update2}
\end{align}
where $v_{k,m}^q$ is the posterior variance of $h_{k,m}$.

To characterize the sparsity of the channel matrix, this paper adopts the spike and slab distribution~\cite{{Lin_CL'17}} to model the a priori distribution of ${\bf H}$, which can be expressed as
\begin{equation}\label{Eq:Pri_Spike_Slab}
\begin{aligned}
\!\!\! p_0\left({\bf H}\right) &= \prod\limits_{m = 1}^M{\prod\limits_{k = 1}^K{ p_0\left(h_{k,m}\right) }}\\
                               &= \prod\limits_{m = 1}^M{\prod\limits_{k = 1}^K\left[ \left(1 - \gamma_{k,m}\right)\delta\left(h_{k,m}\right) + {\gamma_{k,m}}f\left(h_{k,m}\right) \right]}, \!\!\!
\end{aligned}
\end{equation}
where $0 < \gamma_{k,m} <1$ is the sparsity ratio, i.e., the probability of $h_{k,m}$ being non-zero, $\delta\left(\cdot\right)$ is the Dirac delta function.
The a priori distribution of channel gains $f\left(h_{k,m}\right)$ is related to the channel model ${\bf h}_{b,k} = \rho_{b,k}{\widetilde {\bf h}_{b,k}}$, where $b = \lceil m/M_c \rceil$ and ${\widetilde {\bf h}_{b,k}}$ is given in (\ref{Eq:Ch_Model}).
Furthermore, this paper adopts the one-ring channel model, where the UEs are located in a local rich scattering environment, i.e., the number of MPCs $L_{b,k}$ can be large but the angular spread can be limited.
Hence, given $\beta_{b,k}^l \sim {\cal CN}\left( \beta_{b,k}^l; 0,1 \right)$, we assume $f\left(h_{k,m}\right) = {\cal CN}\left(h_{k,m};\mu_{k,m},\tau_{k,m}\right)$ according to the central limit theorem, where
\begin{equation}\label{Eq:Ch_Mean}
\begin{aligned}
\mu_{k,m} &= \mathbb{E}\left[\sqrt{P_k}\rho_{b,k} \sum_{l = 1}^{L_{b,k}} \beta_{b,k}^l e^{-j2{\pi}(m-bM_c)\phi_{b,k}^l} e^{-j2\pi\tau_{b,k}^lf} \right]\\
          &= 0,
\end{aligned}
\end{equation}
and
\begin{equation}\label{Eq:Ch_Var}
\begin{aligned}
\tau_{k,m}\! &=\! \mathbb{E}\left[P_k\rho_{b,k}^2\left(\sum_{l = 1}^{L_{b,k}} \beta_{b,k}^l e^{-j2{\pi}(m-bM_c)\phi_{b,k}^l} e^{-j2\pi\tau_{b,k}^lf}\right)^2 \right]\\
                  &- \mu_{k,m}^2 = P_k\rho_{b,k}^2L_{b,k}.
\end{aligned}
\end{equation}
By exploiting this a priori model in (\ref{Eq:Post_Approx1}), the posterior distribution of $h_{k,m}$ is obtained as follows
\begin{equation}\label{Eq:Post_Approx2}
\begin{aligned}
\!\!\!p\left(h_{k,m}|A_{k,m}^q, B_{k,m}^q\right) &= \left(1 - \theta_{k,m}^q\right)\delta\left(h_{k,m}\right)\\
                                           &+ \theta_{k,m}^q{\cal CN}\left(h_{k,m}; Z_{k,m}^q, V_{k,m}^q\right), \!\!\!
\end{aligned}
\end{equation}
where
\begin{align}
Z_{k,m}^q &= \frac{\tau_{k,m}A_{k,m}^q + \mu_{k,m}B_{k,m}^q} {B_{k,m}^q + \tau_{k,m}}, \label{Eq:Post_A} \\
V_{k,m}^q &= \frac{\tau_{k,m}B_{k,m}^q} {\tau_{k,m} + B_{k,m}^q},\\
{\cal J}_{k,m}^q &= \ln{\frac{B_{k,m}^q}{B_{k,m}^q + \tau_{k,m}}} + \frac{\left|A_{k,m}^q\right|^2}{B_{k,m}^q} - \frac{\left|A_{k,m}^q - \mu_{k,m}\right|^2}{\left(B_{k,m}^q + \tau_{k,m}\right)},\\
\theta_{k,m}^q &= \frac{\gamma_{k,m}}{\gamma_{k,m} + \left(1 - \gamma_{k,m}\right)\exp\left( - {\cal J}_{k,m}^q\right)},
\end{align}
and $\theta_{k,m}^q$ is referred to as the \emph{belief indicator}.
The posterior mean and variance of $h_{k,m}$ can now be explicitly calculated as
\begin{align}
{\hat h}_{k,m} &= {\theta_{k,m}^q}{Z_{k,m}^q}, \label{Eq:Post_Mean} \\
v_{k,m}^q &= {\theta_{k,m}^q}\left(\left|Z_{k,m}^q\right|^2 + V_{k,m}^q\right) - \left|{\hat h}_{k,m}\right|^2, \label{Eq:Post_Var}
\end{align}
respectively.
The equations (\ref{Eq:Var_Update1})-(\ref{Eq:Fac_Update2}) and (\ref{Eq:Post_A})-(\ref{Eq:Post_Var}) make up the key steps of the basic AMP algorithm, which provides a simplified approach to calculate the MMSE estimate of ${\bf H}$.
Here, we assume the CPU can acquire the full knowledge of the sparsity ratio $\gamma_{k,m}$ and the noise variance ${\widehat \sigma}$, which is an impractical assumption.
The reason is that, for practical cell-free massive MIMO systems, the varying numbers of active UEs leads to the varying channel sparsity level $\gamma_{k,m}$.
Moreover, when performing angular-domain CE, the variance of the effective noise ${\widetilde {\bf N}}_p$ is hard to compute as the estimation error of AUD would be unknown.
For facilitating the practical implementation of the algorithm, the EM is employed to learn the unknown hyper-parameters,
\begin{align}
{\widehat \sigma^{q+1}} &= \frac{1}{GM} \! \sum\nolimits_{g=1}^G \sum\nolimits_{m=1}^M \!\! {\left[\frac{\left|{\widehat y_{g,m}} \!-\! D_{g,m}^q\right|^2}{\left|1 \!+\! C_{g,m}^q/{\widehat \sigma^q}\right|^2} + \frac{{{\widehat \sigma^q}}{C_{g,m}^q}}{{\widehat \sigma^q} \!+\! C_{g,m}^q}\right]}, \label{Eq:Noi_Var} \\
\gamma_{k,m}^{q+1} &= \theta_{k,m}^{q+1} = \frac{\gamma_{k,m}^q}{\gamma_{k,m}^q + (1-\gamma_{k,m}^q)\exp\left(-{\cal J}_{k,m}^q\right)}. \label{Eq:Sparse_Ratio}
\end{align}
Finally, given the posterior mean and variance of the channel matrix, the extrinsic messages of SLM module are given as~\cite{Meng_SPL'18}
\begin{align}
{\bf Y}^{\rm pri} &= {\bf S}{\widehat {\bf H}} + \frac{{\bf C}^q}{{\widehat \sigma} + {\bf C}^{q-1}}\circ\left({\widehat {\bf Y}} - {\bf D}^{q-1}\right),\label{Eq:ext_slm1}\\
V^{\rm pri} &= \frac{1}{GM}\|{\bf C}\|_{\rm F}^2,\label{Eq:ext_slm2}
\end{align}
where $\circ$ denotes the Hadamard product.

Next, we extend the key steps of the generalized AMP algorithm derived above, i.e., (\ref{Eq:quan_post_mean})-(\ref{Eq:ext_mean_nonlin}), (\ref{Eq:Var_Update1})-(\ref{Eq:Fac_Update2}), (\ref{Eq:Post_A})-(\ref{Eq:ext_slm2}), to the multiple subcarriers case, where the spatial-domain or angular-domain structured sparsity of the channel matrix is exploited to enhance the CS recovery performance.
The resulted algorithm is referred to as the SS-GAMP algorithm, which is summarized in \emph{Algorithm \ref{Alg:SS-GAMP}}.
Specifically, in \emph{lines \ref{Step:quan_post}-\ref{Step:Hyper_Update}}, the messages are updated independently for all subcarriers.
Moreover, as the matrix estimation problem is decoupled into multiple scalar estimation problems, as shown in (\ref{Eq:AMP_Decouple}), the variables for calculating the associated messages are also computed independently for all $b, k, g$, and $m$.
\emph{Line \ref{Step:Damp1}} employs a damping parameter $\rho = 0.3$ to prevent the SS-GAMP algorithm from diverging \cite{Rangan_ISIT'14}.
Note that except for \emph{line \ref{Step:Hyper_Refine}}, all variables in the SS-GAMP algorithm are updated independently.
In \emph{line \ref{Step:Hyper_Refine}}, the sparsity ratio $\gamma_{p,b,k,m}^{q+1}$ associated with different $p$, $b$, and $m$, are jointly refined based on the spatial-domain or angular-domain structured sparsity of the channel matrix.

\begin{algorithm}[t]
\caption{SS-GAMP Algorithm}
\label{Alg:SS-GAMP}
\begin{algorithmic}[1]
\REQUIRE $\forall p,b$\;: Quantized received signals ${\overline {\bf Y}_{p,b}}$, ${\bf S}_p$, $\Delta_b$, $y_b^{\rm max}$, and $y_b^{\rm min}$; $\rho$, the maximum numbers of AMP and turbo iterations, $T_{\rm amp}$ and $T_{\rm tur}$, and the termination threshold $\eta$.
\ENSURE $\forall p,b,k,m$\;: Estimated channel matrices $\{ {\widehat {\bf H}_{p,b}} \}_{p=1}^P$ and the related belief indicators $\theta_{p,b,k,m}$. $\%$ In the reminder, $\forall m$ denotes $\forall m \in [M_c]$\\
\STATE $\forall p,b,k,m,g$: Set AMP iteration index $q$ to 1, set turbo iteration index $i$ to 1, initialize the $\gamma_{p,b,k,m}$ and ${\widehat \sigma}$ as in \cite{Ke_TSP'20}, and initialize other parameters as $y_{p,b,g,m}^{\rm pri}(1) = 0$, $V^{\rm pri}(1) = 10^6$, $C_{p,b,g,m}^0 = 1$, $D_{p,b,g,m}^0 = y_{p,b,g,m}$, $\hat h_{p,b,k,m}^1 = \mu_{p,b,k,m}^1$, $v_{p,b,k,m}^1 = \tau_{p,b,k,m}^1$.\\
\label{Step:Initial}
\FOR {$i \le T_{\rm tur}$}
\STATE $\forall p,b$: Compute the posterior mean ${\bf Y}_{p,b}^{\rm post}(i)$ and posterior variance $V_{p,b}^{\rm post}(i)$ of the un-quantized received signals ${\bf Y}_{p,b}$, as in (\ref{Eq:quan_post_mean})-(\ref{Eq:eta_2}).
\label{Step:quan_post}
\STATE $\forall p,b$\;: Compute the extrinsic messages of nonlinear module, ${\widehat {\bf Y}}_{p,b}(i)$ and ${\widehat \sigma}_{p,b}(i)$, as in (\ref{Eq:ext_mean1}) and (\ref{Eq:ext_mean_nonlin}), respectively, and ${\widehat \sigma}(i) = \frac{1}{PB}{\widehat \sigma_{p,b}}(i)$.
\REPEAT
\STATE $\forall p,b,g,m$: Update $C_{p,b,g,m}^q$ and $D_{p,b,g,m}^q$ according to (\ref{Eq:Fac_Update1}) and (\ref{Eq:Fac_Update2}).
\label{Step:Fac_Update}
\STATE $C_{p,b,g,m}^q = {\rho}{C_{p,b,g,m}^{q-1}} + \left(1-\rho\right){C_{p,b,g,m}^q}$,\; $D_{p,b,g,m}^q = {\rho}{D_{p,b,g,m}^{q-1}} + \left(1-\rho\right){D_{p,b,g,m}^q}$.
\label{Step:Damp1}
\STATE $\forall p,b,k,m$: Update $B_{p,b,k,m}^q$ and $A_{p,b,k,m}^q$ according to (\ref{Eq:Var_Update1}) and (\ref{Eq:Var_Update2}).
\label{Step:Var_Update}
\STATE $\forall p,b,k,m$: Compute posterior mean $\hat h_{p,b,k,m}^{q+1}$ and posterior variance $v_{p,b,k,m}^{q+1}$ of the cannel matrix according to (\ref{Eq:Post_Mean}) and (\ref{Eq:Post_Var}), respectively.
\label{Step:Post__Mean_Var}
\STATE $\forall p,b,k,m$: Update the $\gamma_{p,b,k,m}^{q+1}$ as in (\ref{Eq:Sparse_Ratio}). Moreover, ${\widehat \sigma^{q+1}}\! =\! \frac{\sum_p\sum_b {\widehat \sigma_{p,b}^{q+1}}}{PB}$, where ${\widehat \sigma_{p,b}^{q+1}}$ is given in (\ref{Eq:Noi_Var}).\!\!\!\!\!
\label{Step:Hyper_Update}
\STATE $\forall p,b,k,m$: Refine the update rule of the sparsity ratio $\gamma_{p,b,k,m}^{q+1}$ based on the structured sparsity of the channel matrix, as in (\ref{Eq:Refine_1})-(\ref{Eq:Refine_Ang}).
\label{Step:Hyper_Refine}
\STATE $q = q + 1$.
\UNTIL $q\! \ge \!T_{\rm amp}$ or $\sum_p\!{\left\|{\widehat {\bf H}}_p^q\! -\! {\widehat {\bf H}}_p^{q-1}\right\|_{\rm F}^2} / \sum_p\!{\left\|{\widehat {\bf H}}_p^{q-1}\right\|_{\rm F}^2}\! <\! \eta$.
\STATE $i = i + 1$.
\STATE Compute the extrinsic messages of SLM module as in (\ref{Eq:ext_slm1}) and (\ref{Eq:ext_slm2}).
\ENDFOR
\RETURN $\{{\widehat {\bf H}_{p,b}^{q-1}}\}_{p=1}^P, \forall b$;\; ${\theta_{p,b,k,m}} = \gamma_{p,b,k,m}^{q-1}$, $\forall p,b,k,m$.
\end{algorithmic}
\end{algorithm}

When applying the SS-GAMP algorithm to the spatial-domain channel model (\ref{Eq:Cloud_Process_Spa}) for AUD, the spatial-domain structured sparsity of $\{ {\bf H}_{p} \}_{p=1}^P$ is considered.
For the channel matrix between all UEs to a specific AP $b$, the channel vectors $\left[{\bf H}_{p,b}\right]_{:,m}$ observed at different pilot subcarriers and different AP antennas have a common sparsity, as described in (\ref{Eq:Sparse_SPa_1})-(\ref{Eq:Sparse_SPa_3}) and illustrated in Fig. \ref{Fig4}(a).
Meanwhile, the sparsity ratio $\gamma_{p,b,k,m}$ is the probability that the $(k,m)$-th element of $\left[{\bf H}_{p,b}\right]_{k,m}$ is non-zero.
Hence, for channel matrices $\{ {\bf H}_{p,b} \}_{p=1}^P$, the elements associated with the same UE share a common sparsity ratio, so we consider
\begin{equation}\label{Eq:Refine_1}
{\widetilde \gamma}_{b,k}^{q+1} = \frac{1}{\left|{\cal N}_{p,b,k,m}\right|_c} \sum\nolimits_{(o,b,k,u) \in {\cal N}_{p,b,k,m}} \theta_{o,b,k,u}^{q+1},
\end{equation}
where
\begin{equation} \label{Eq:NNS_1}
{\cal N}_{p,b,k,m} = \left\{\left(o, b, k, u\right)|o\! =\! 1, \cdots, P; \; u\! =\! 1, \cdots, M_c\right\}.\!\!\!\!
\end{equation}
Additionally, by further considering the approximate common sparsity between channel matrices $\{ {\bf H}_{p,b} \}_{p=1}^P$ for different $b$, the update rule for the sparsity ratio can be finally refined as
\begin{equation} \label{Eq:Refine_2}
{\gamma}_{p,b,k,m}^{q+1} = {\gamma}_k^{q+1} = \sum\nolimits_{b=1}^B \frac{1} {{d_{b,k}}\Delta_k}{\widetilde \gamma}_{b,k}^{q+1},
\end{equation}
where $d_{b,k}$ denotes the distance between the $k$-th UE and the $b$-th AP, and $\Delta_k = \sum\nolimits_{b=1}^B 1/ {{d_{b,k}}}$.
We can explain $(\ref{Eq:Refine_2})$ intuitively from a UE-centric perspective.
Specifically, for a specific UE $k$, its activity observed from the adjacent APs can be more reliable than that observed from the remote APs.
Therefore, we consider a weighted method to refine the sparsity ratio, i.e., compared to the remote APs, the adjacent APs contribute more weights to the value of ${\gamma}_k^{q+1}$.

Due to the DFT transformation in (\ref{Eq:Rb_Ang_1}), the angular domain received signals ${\bf R}_{p,b}, \forall p, b$ are not consistent with the quantization codebook of the corresponding GLM, i.e., $r_{p,b,g,m} \notin {\cal C}_b, \forall p, b, g, m$.
This will lead to an unreliable estimate of ${\bf Y}_{p,b}$, as the nonlinear module of SS-GAMP algorithm is developed based on the quantization codebook.
Hence, for CE, we directly apply the SLM module of SS-GAMP algorithm to (\ref{Eq:Cloud_Process_Ang}), where the quantization error is treated as noise.
Here, the virtual-angular domain sparsity of massive MIMO channels is further taken into account.
However, this angular-domain sparsity destroys the structured sparsity over different AP antennas.
Hence, the clustered sparsity illustrated in Fig. \ref{Fig4}(b) is leveraged to refine the sparsity ratio.
Specifically, define the neighbors of $w_{p,b,k,m}$ as
\begin{equation}\label{Eq:NNS_2}
\begin{aligned}
{\widetilde {\cal N}}_{p,b,k,m} = &\left\{\left(p - 1, b, k, m\right), \left(p + 1, b, k, m\right), \right.\\
                                  &\left. \left(p, b, k, m - 1\right), \left(p, b, k, m + 1\right) \right\},
\end{aligned}
\end{equation}
$w_{p,b,k,m}$ and the elements of ${\widetilde {\cal N}}_{p,b,k,m}$ tend to be simultaneously either zero or non-zero, and the update rule of ${\gamma}_{p,b,k,m}$ is given as
\begin{equation} \label{Eq:Refine_Ang}
{\gamma}_{p,b,k,m}^{q+1} = \frac{1}{\left|{\widetilde {\cal N}}_{p,b,k,m}\right|_c} \sum\nolimits_{(o,b,k,u) \in {\widetilde {\cal N}}_{p,b,k,m}} \theta_{o,b,k,u}^{q+1}.
\end{equation}

\begin{remark}
In contrast to the conventional CS algorithms for SLM, the proposed SS-GAMP mainly shows its superiority in low-resolution quantization cases.
When the quantization accuracy is good enough, i.e., the number of quantization bits $Q$ is large, the quantization error can be negligible.
In this case, we can directly apply the SLM module of SS-GAMP algorithm to (\ref{Eq:Cloud_Process_Spa}) for AUD, which can reduce the computational complexity with negligible performance loss.
\end{remark}

\subsection{SIC-Based AUD and CE Algorithm}
\label{Sec:IV-B}

The AUD and CE can be jointly realized by applying the SS-GAMP algorithm to (\ref{Eq:Cloud_Process_Spa}) or (\ref{Eq:Cloud_Process_Ang}).
However, these solutions can not fully exploit the enhanced sparsity of $\left\{{\bf W}_p\right\}_{p = 1}^P$ and the structured sparsity of $\left\{{\bf H}_p\right\}_{p = 1}^P$.
In this section, based on the SS-GAMP algorithm, we develop a SIC-based AUD and CE algorithm for alternately detecting active UEs based on (\ref{Eq:Cloud_Process_Spa}) and estimating their channels based on (\ref{Eq:Cloud_Process_Ang}), so that the massive access performance can be further improved.

\begin{figure}[t]
    \captionsetup{font={footnotesize}, name = {Fig.}, labelsep = period}
    \centering
    \includegraphics[width=0.8\columnwidth,keepaspectratio]
    {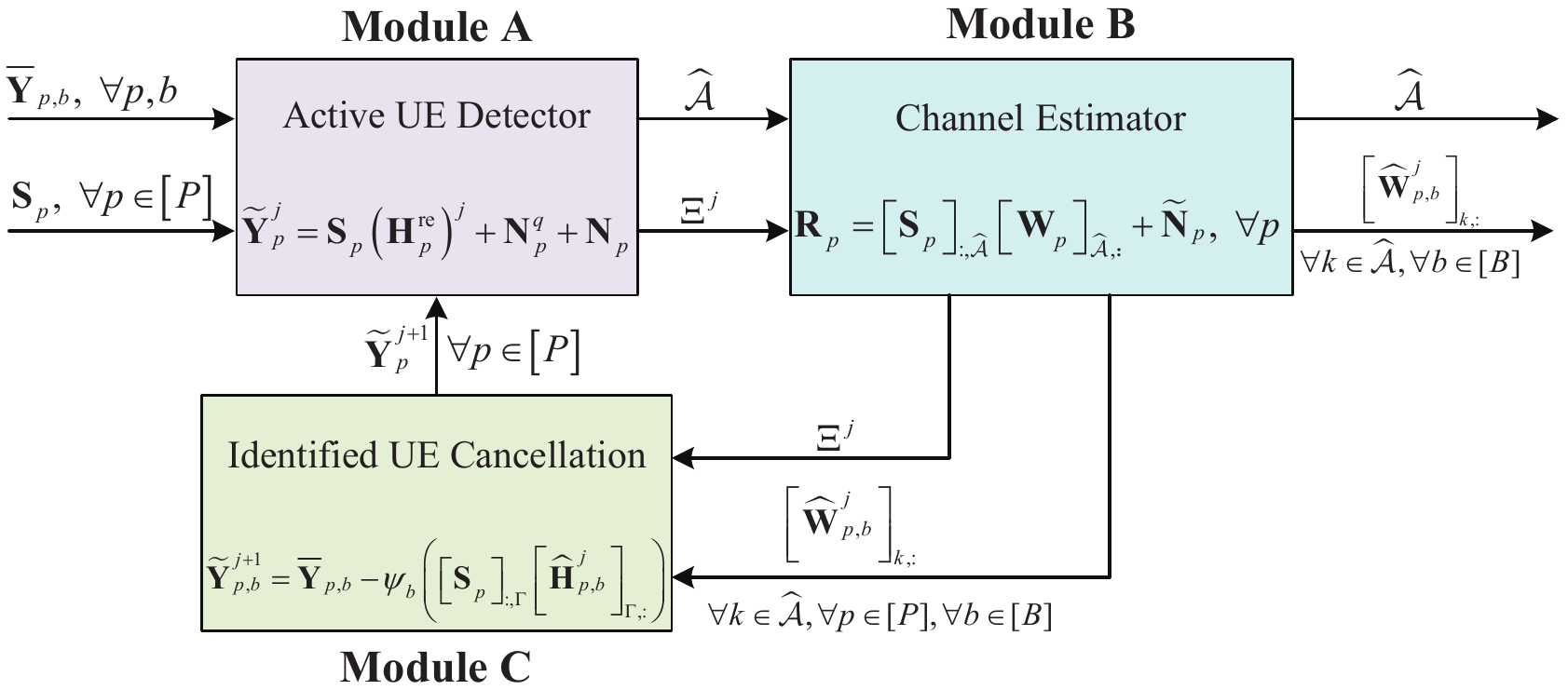}
	\caption{Block diagram of the proposed SIC-based AUD and CE algorithm.}
    \label{Fig:6}
\end{figure}

The procedure of the proposed algorithm is summarized in \emph{Algorithm \ref{Alg:SIC_AUD_CE}} and is illustrated in Fig.~\ref{Fig:6}, which mainly consists of three modules.
Specifically, in each SIC iteration, the spatial-domain active UE detector (module A) acquires a rough AUS estimate and a relatively reliable AUS estimate (i.e., ${\widehat {\cal A}}$ and $\Xi^j$, respectively, in Fig.~\ref{Fig:6}), which are passed to the angular-domain channel estimator (module B); subsequently, module B estimates the channels of the identified active UEs in ${\widehat {\cal A}}$; finally, based on the AUS and CSI estimates, module C updates the residual received signals by cancelling the components associated with the active UEs identified in $\Xi^j$, and the residual received signals are passed to module A.
The three modules are executed alternately in an iterative manner until convergence.
Next, we will detail three modules as follow.

\emph{Module A: Spatial-domain AUD.}
In the first SIC iteration ($j = 1$), module A detects active UEs based on the spatial-domain channel model.
Specifically, the SS-GAMP algorithm is applied to model (\ref{Eq:Cloud_Process_Spa}) to acquire the belief indicators $\theta_{p,b,k,m}^j$, $\forall p,b,k,m$, based on which the AUS estimate, denoted as ${\widehat {\cal A}}$, is determined.
It has been proved in \cite{Ke_TSP'20} that if a reliable estimate of $\{ {\bf H}_{b} \}_{p=1}^P$ is acquired after the convergence of the SLM module of SS-GAMP algorithm, belief indicator $\theta_{p,b,k,m}^j$ tends to 1 for $h_{p,b,k,m} \ne 0$ and 0 for $h_{p,b,k,m} = 0$.
Hence, we design a belief indicator-based active UE (BI-AUE) detector as follows
\begin{equation}\label{Eq:BI-AD}
{\widehat \alpha_k} = \left\{ \begin{array}{*{20}{c}}
1, &\frac{1} {PM_c} \sum_{p=1}^P{\sum_{m=1}^{M_c}{\theta_{p,b^*,k,m}^j \ge p_{\rm th}}},\\
0, &\frac{1} {PM_c} \sum_{p=1}^P{\sum_{m=1}^{M_c}{\theta_{p,b^*,k,m}^j < p_{\rm th}}}.
\end{array} \right.
\end{equation}
Here, $k \in [K]$, $b^* = \{b|{\rm min}(d_{b,k}), \forall b\}$, i.e., the $k$-th UE's activity is mainly dependent on the belief indicators $\theta_{p,b^*,k,m}^j$ inferred from the pilot signals received at the $b^*$-th AP, which has the shortest spatial distance (also has the smallest path loss) with the $k$-th UE.
For facilitating the subsequent CE and SIC processing, we utilize the BI-AUE detector to acquire two AUS estimates having different reliability: a rough AUS estimate ${\widehat {\cal A}}$ based on a lower threshold $p_{\rm th} = p_{\rm det}$ and a relatively reliable AUS estimate $\Xi^j$ based on a higher threshold $p_{\rm th} = p_{\rm rel}$, as shown in \emph{lines \ref{Step:AUD_Start}-\ref{Step:AUD_End}} in \emph{Algorithm \ref{Alg:SIC_AUD_CE}}, so that $\Xi^j \subseteq {\widehat {\cal A}}$.
Here, we set $p_{\rm det}$ to 0.1 to reduce the missed detection probability, and $p_{\rm rel}$ is set to 0.9 to guarantee the UEs in $\Xi^j$ are active with high probability.
These two AUS estimates, ${\widehat {\cal A}}$ and $\Xi^j$, are passed to module B.

\begin{algorithm}[t]
\caption{SIC-Based AUD and CE Algorithm}
\label{Alg:SIC_AUD_CE}
\begin{algorithmic}[1]
\REQUIRE $\forall p,b$\;: Quantized received signals ${\overline {\bf Y}_{p,b}}$, ${\bf S}_p$; the number of SIC iterations $T_{\rm sic}$.
\ENSURE $\forall p,b$\;: The AUS estimate ${\widehat {\cal A}}$ and the related CSI estimates $\{\widehat {\bf h}_{p,b,k}\}_{p=1}^P, \forall k \in {\widehat {\cal A}}$.
\STATE Initialization: $j = 1$, $\Xi^0 = \emptyset$, ${\widetilde {\bf Y}}_p^1 = {\overline {\bf Y}_p}$.
\REPEAT
\STATE $k = 0$, ${\widehat {\cal A}} = \Gamma = \emptyset$.\\
\label{Step:Iter_Start}
%// Module A: Spatial-domain channel model based active UE detector
\label{Step:Module_A}
\STATE $\forall p,b,k,m$\;: Acquire $\theta_{p,b,k,m}^j$ by applying the SS-GAMP algorithm to model (\ref{Eq:AUD_Re}).
\label{Step:Spa_AMP}
\FOR {$k \le K$}
\label{Step:AUD_Start}
\IF {$\frac{1}{PM_c}\sum_{p=1}^P{\sum_{m=1}^{M_c}{\theta_{p,b^*,k,m}^j \ge p_{\rm det}}}$}
\STATE ${\widehat {\cal A}}  = {\widehat {\cal A}} \cup \Xi^{j-1} \cup \left\{k\right\}$.\\
\ENDIF
\IF {$\frac{1}{PM_c}\sum_{p=1}^P{\sum_{m=1}^{M_c}{\theta_{p,b^*,k,m}^j \ge p_{\rm rel}}}$}
\STATE $\Xi^j  = \Xi^{j-1}  \cup \left\{k\right\}$.\\
\ENDIF
\STATE $k = k + 1$.
\ENDFOR\; $\%$ Here, $b^* = \{b|{\rm min}(d_{b,k}), \forall b \in [B]\}$.\\
\label{Step:AUD_End}
%// Module B: Angular-domain channel estimation for identified active UEs
\STATE $\forall p$\;: ${\bf R}_p = \left[{\overline {\bf Y}_{p,1}}{\bf A}_R^*, {\overline {\bf Y}_{p,2}}{\bf A}_R^*, \cdots, {\overline {\bf Y}_{p,B}}{\bf A}_R^*  \right]$, ${\widehat {\bf W}_p}^j = {\bf 0}_{K \times M}$.
\STATE $\forall p,b$\;: Acquire the channel vectors $\left[{\widehat {\bf W}}_{p,b}^j\right]_{k,:}, \forall k\! \in\! {\widehat {\cal A}}$, by applying the SLM module of SS-GAMP algorithm to model (\ref{Eq:CE_Act}).\\
\label{Step:Ang_AMP}
%// Module C: Identified UE cancellation
\STATE Acquire set $\Gamma$, $\Gamma \subseteq \Xi^j$, and $\left|\Gamma\right|_c / \left|\Xi^j\right|_c = \lambda_{\rm aus}$. \% The elements in $\Gamma$ are randomly selected from $\Xi^j$.
\label{Step:Re_Set}
\STATE $\forall p$\; : ${\widehat {\bf H}_p^j} = \left[ {\widehat {\bf W}_{p,1}^j}{\bf A}_R^{\rm T}, {\widehat {\bf W}_{p,2}^j}{\bf A}_R^{\rm T}, \cdots, {\widehat {\bf W}_{p,B}^j}{\bf A}_R^{\rm T} \right]$.
\STATE $\forall p,b$\; : ${\widetilde {\bf Y}}_{p,b}^{j+1} = {\overline {\bf Y}_{p,b}} - \psi_b\left(\left[{\bf S}_p\right]_{:,\Gamma}\left[ \widehat {\bf H}_{p,b}^j\right]_{\Gamma,:}\right)$. $\%$  $\psi_b(\cdot)$ is applied only when the low-resolution quantization is considered.
\label{Step:Compute_Re}
\STATE $j = j + 1$.
\label{Step:Iter_End}
\UNTIL $j\! > \!T_{\rm tur}$.
\RETURN $\forall p,b$\;: ${\widehat {\cal A}}$;\; ${\widehat {\bf h}}_{p,b,k} = \left[ {\widehat {\bf H}_{p,b}^{j-1}} \right]_{k,:}, \forall k \in {\widehat {\cal A}}$.
\end{algorithmic}
\end{algorithm}

\emph{2) Module B: Angular-domain CE for identified active UEs.}
In module B, given the rough AUS estimate ${\widehat {\cal A}}$, the angular-domain channel vectors of the UEs in ${\widehat {\cal A}}$, i.e., $\left[{\bf W}_p\right]_{{\widehat {\cal A}},:}$, are estimated based on the model in (\ref{Eq:Cloud_Process_Ang}) as follows
\begin{equation}\label{Eq:CE_Act}
{\bf R}_p = \left[{\bf S}_p\right]_{:,{\widehat {\cal A}}}\left[{\bf W}_p\right]_{{\widehat {\cal A}},:} + {\widetilde {\bf N} _p},\; \forall p \in \left[P\right],
\end{equation}
where $\left[{\bf S}_p\right]_{:,{\widehat {\cal A}}} \in \mathbb{C}^{G \times \left|{\widehat {\cal A}}\right|_c}$ and $\left[{\bf W}_p\right]_{{\widehat {\cal A}},:} \in \mathbb{C}^{\left|{\widehat {\cal A}}\right|_c \times M}$ are sub-matrices of ${\bf S}_p$ and ${\bf W}_p$, respectively, ${\widetilde {\bf N}_p} = \left[{\bf S}_p\right]_{:,{{\cal K} - {\widehat {\cal A}}}}\left[{\bf W}_p\right]_{{{\cal K} - {\widehat {\cal A}}},:} + {\overline {\bf N}_p^q} + {\overline {\bf N}}_p$, ${\cal K}$ is the set of all potential UEs, and ${\cal K} - {\widehat {\cal A}}$ denotes the difference set of sets $\cal K$ and ${\widehat {\cal A}}$.
Note that ${\widetilde {\bf N}_p}$ is the effective noise including AWGN, quantization error, and the estimation error of AUD.
Furthermore, if ${\cal A} \subseteq {\widehat {\cal A}}$, we have ${\widetilde {\bf N}_p} = {\overline {\bf N}_p^q} + {\overline {\bf N}}_p$.
Hence, to reduce the power of ${\widetilde {\bf N}_p}$, a low missed detection probability is desirable.
According to the angular-domain structured sparsity of $\{ {\bf W}_{p} \}_{p=1}^P$, as described in (\ref{Eq:Sparse_Ang_1})-(\ref{Eq:Sparse_Ang_3}) and illustrated in Fig. \ref{Fig4}(b), the low-dimensional channel matrix $\left[{\bf W}_p\right]_{{\widehat {\cal A}},:}$ is still sparse.
Hence, we can estimate $\left[{\bf W}_p\right]_{{\widehat {\cal A}},:}, \forall p$, by applying the SLM module of SS-GAMP algorithm to (\ref{Eq:CE_Act}), see \emph{line \ref{Step:Ang_AMP}} of \emph{Algorithm \ref{Alg:SIC_AUD_CE}}.
Finally, the reliable AUS estimate $\Xi^j$ and the related CSI estimate are passed to module C.

\emph{3) Module C: Identified UE cancellation.}
Since the active UEs in $\Xi^j$ are reliably detected in module A and their CSI is estimated in module B, the signals received from the UEs in $\Gamma$, a subset of $\Xi^j$, are removed from ${\overline {\bf Y}_p}$ to enhance the sparsity of the channel matrix for AUD in the next SIC iteration.
The residual received signals ${\widetilde {\bf Y}}_p^j, \forall p$ are computed in \emph{lines \ref{Step:Re_Set}}-\emph{\ref{Step:Compute_Re}}, and are passed to module A.
In the following SIC iterations ($j > 1$), the AUD problem in module A is to recover $\left({\bf H}_p^{\rm re}\right)^j$  based on the following model
\begin{equation}\label{Eq:AUD_Re}
{\widetilde {\bf Y}}_p^j = {\bf S}_p\left({\bf H}_p^{\rm re}\right)^j + {\bf N}_p^q + {\bf N}_p,  \forall p \in \left[P\right],
\end{equation}
where ${\widetilde {\bf Y}}_p^j$ denotes the residual received signals in the $j$-th SIC iteration, $\left({\bf H}_p^{\rm re}\right)^j = {\bf H}_p - {\widetilde {\bf H}_p^j}$, and ${\widetilde {\bf H}_p^j} \in \mathbb{C}^{K \times M}$ is defined as $[{\widetilde {\bf H}_p}^j]_{\Gamma,:} = [\widehat {\bf H}_p^{j-1}]_{\Gamma,:}$, while $[{\widetilde {\bf H}_p^j}]_{{\cal K}-\Gamma,:} = {\bf 0}_{\left|{\cal K}-\Gamma\right|_c \times M}$.
To guarantee the robustness of the SS-GAMP-based AUD, we only remove the signals received from a part of the UEs in $\Xi^j$, i.e., $\lambda_{\rm aus} < 1$ (e.g., we consider $\lambda_{\rm aus} = 0.8$).

Modules A, B, and C will be executed alternately in an iterative manner.
Since the $\left({\bf H}_p^{\rm re}\right)^j$ becomes sparser and the CSI estimates of the UEs in ${\widehat {\cal A}}$ are iteratively re-estimated as the SIC iterations proceed, the ${\widehat {\cal A}}$ and the corresponding CSI estimates are constantly refined.
Therefore, compared to the joint AUD and CE solutions without SIC, the proposed SIC-based scheme facilitates more reliable AUD and CE with a significant reduction in access latency.
However, as the SS-GAMP algorithm is called twice in each SIC iteration and the identified UE cancellation requires additional matrix multiplication, the performance improvement is at the cost of a higher computational complexity.

\section{Differences Between Cloud Computing and Edge Computing Paradigms}
\label{Sec:V}

This section compares cloud computing and edge computing in terms of their algorithm implementation, computational complexity, access latency, and the cost of AP deployment.
Compared with cloud computing, edge computing has the advantages of alleviating the burden on backhaul links and CPU, a faster access response, and supporting more flexible AP cooperation, while increases the cost of large-scale AP deployment.

\subsection{Algorithm Implementation}
\label{Sec:V-A}

For cloud computing paradigm, the detailed procedure of the proposed AUD and CE approach is summarized in \emph{Algorithms \ref{Alg:SS-GAMP}} and \emph{\ref{Alg:SIC_AUD_CE}}.
It is clear that the signals collected from $B$ APs are processed in parallel in \emph{lines \ref{Step:quan_post}-\ref{Step:Hyper_Update}} of \emph{Algorithm \ref{Alg:SS-GAMP}}, and are centrally processed in \emph{line \ref{Step:Hyper_Refine}} only.
Intuitively, \emph{line \ref{Step:Hyper_Refine}} leverages the structured sparsity described in Section \ref{Sec:II-C} to refine the update rule of the sparsity ratio $\gamma_{p,b,k,m}$, as in (\ref{Eq:Refine_1})-(\ref{Eq:Refine_Ang}).
Note that this paper considers a large-scale network to serve a vast area, so the channel strength from a specific active UE to far away APs approximates zero due to the large-scale fading caused by severe path loss.
Hence, for this specific UE, the signals received at the remote APs have a negligible effect on refining the sparsity ratio.
This reveals that centrally processing all the APs' received signals at the CPU for jointly refining sparsity ratio maybe not an efficient way.

While for edge computing paradigm, the SIC-based AUD and CE algorithm summarized in \emph{Algorithm \ref{Alg:SIC_AUD_CE}} can be directly applied based on (\ref{Eq:Edge_Spa}) and (\ref{Eq:Edge_Ang}) for detecting active UEs and estimating their channels, respectively.
Here, the AUD and CE problems for the whole network are locally processed at multiple DPU-APs in close proximity to the UEs.
Clearly, the APs in the edge computing paradigm are divided into several groups, and each group seeks to detect only part of the total UEs.
We can also explain the AP and UE association from a UE-centric perspective, that is to say, for a specific UE, its activity and CSI can be estimated by jointly processing the signals received at its nearest one DPU-AP and $(N_{co}-1)$ APs.
Compared to cloud computing, the edge computing enables more flexible AP cooperation by considering different numbers of cooperative APs and reduces the transmission burden on backhaul links.

\subsection{Computational Complexity}
\label{sec:V-B}

For each SIC iteration in cloud computing, the complexity\footnote{Here, we mainly focus on the maximum number of required complex multiplications.} of SS-GAMP algorithm is in order of ${\cal O}\left(T_{\rm amp}\left(4GKMP + 3GKP + 16GMP + 20KMP\right) + T_{\rm tur}\right.$ $\times \left.\left(GKMP+GMP\right)\right)$, the complexity of DFT is ${\cal O}\left(2BM_c^2P\right)$, and the complexity of computing residual received signal for SIC is ${\cal O}\left(GK_{\rm sic}MP\right)$, $K_{\rm sic}$ is the number of UEs for cancellation.
Hence, the overall complexity of the processing tasks at CPU is given as
\begin{equation}
\begin{aligned}
\!\!C_{\rm cloud}\! =\! {\cal O}(& T_{\rm sic}[2T_{\rm amp}\left(4GKMP\! +\! 3GKP\! +\! 16GMP\right.\\
                         &\left.+ 20KMP\right) + T_{\rm tur}\left(GKMP+GMP\right)\\
                         &+ 2BM_c^2P + GK_aMP]).\!\!
\end{aligned}
\end{equation}

While for edge computing paradigm, the complexity of SIC-based algorithm applied in the $i$-th DPU-AP is
\begin{equation}
\begin{aligned}
C_{\rm edge}^i = {\cal O}(& T_{\rm sic}[2T_{\rm amp}\left(4GK_iM_iP + 3GK_iP + 16GM_iP\right.\\
                          & \left.+ 20K_iM_iP\right) + T_{\rm tur}\left(GKM_iP+GM_iP\right)\\
                          & + 2N_{co}M_c^2P + GK_a^iMP]),
\end{aligned}
\end{equation}
where $K_i$ is the number of UEs detected by the $i$-th DPU-AP, $K_a^i = {\gamma}K_i$, and $M_i = N_{co}M_c$.
Since each DPU-AP seeks to detect only part of the total UEs (i.e., $K_i < K$) and $N_{co} < B$, we have $C_{\rm cloud} > C_{\rm edge}^i$.
Hence, by splitting the signal processing task of the whole network and executing related computations at the edge of the network, edge computing can alleviate the computing burden on CPU.

\subsection{Access Latency}
\label{sec:V-C}

The access latency of grant-free massive access consists of three components: pilot transmission time, propagation latency, and computation latency.
First, the pilot transmission time depends on the adopted frame structure and the pilot length, which are the same for both cloud computing and edge computing.
Second, the DPU-APs in edge computing are deployed at the edge of the network, while the CPU in cloud computing is usually very far away from the UEs.
This results in a much smaller propagation delay for edge computing than that for cloud computing.
Furthermore, cloud computing requires the information to pass through several networks including the radio access network, backhaul network, and core network, where traffic control, routing, and other network-management operations can contribute to excessive delays.
Last, the CPU can have a massive computation capacity than that of DPU.
However, the CPU has to be shared by a large number of other services, and the computational complexity of processing tasks at CPU is much larger than that at DPU, as described in Section~\ref{sec:V-B}.
Moreover, with the rapid development of the processors, the DPU is powerful enough for running highly sophisticated computing programs.
Therefore, the cloud computing and edge computing can have similar computation latencies.
According to the analysis above, edge computing can have a faster access response than cloud computing.

\subsection{Cost of AP Deployment}
\label{sec:V-C}

For cloud computing, all APs are only designed for transmitting and receiving signals, where only antennas and radio frequency chains are needed.
While for edge computing, part of APs should employ extra DPUs so that these APs can be upgraded to DPU-APs.
In cell-free massive MIMO with quantities of APs, this will increase the cost of AP deployment.
Furthermore, edge computing also requires some extra links between APs and DPU-APs.

\section{Simulation Results}
\label{Sec:VI}

This section conducts simulations to validate the superiority of the proposed massive access schemes.
The simulation parameters are provided in Table \ref{Tab:I}.
We consider a typical massive access scenario in cell-free massive MIMO systems, where $K = 2800$ UEs are uniformly distributed in the network and $B = 7$ APs are geographically distributed to serve these UEs.
To reduce the computational complexity, in pilot phase, we only use the signals received at ${\widetilde P}$ out of total $P$ pilot subcarriers for AUD, where ${\widetilde P} \le P$.
With the obtained AUS estimate ${\widehat {\cal A}}$, all $P$ pilot subchannels can be estimated by applying the SLM module of SS-GAMP algorithm to~(\ref{Eq:CE_Act}).

\begin{table}[t]
\caption{Simulation Parameters}
\label{Tab:I}
\center
\begin{tabular}{@{}cc@{}}
\toprule
Parameter                                                      & Value                               \\ \midrule
\vspace{0.2mm}
Radius of the network coverage                                 & 2.65 km                             \\
\vspace{0.2mm}
AP-to-AP distance                                              & $\sqrt{3}$ km                       \\
\vspace{0.2mm}
Number of active UEs $K_a$                                     & 140                                 \\
\vspace{0.2mm}
\vspace{0.2mm}
Transmit power $P_k$                                           & 23 dBm                              \\
\vspace{0.2mm}
Background noise power                                         & -174 dBm/Hz                         \\
\vspace{0.2mm}
OFDM's DFT size $P$ in pilot phase                             & 64                                  \\
\vspace{0.2mm}
OFDM's DFT size $N$ in data phase                              & 2048                                \\
\vspace{0.2mm}
Cyclic prefix length $N_{CP}$                                  & 64                                  \\
\vspace{0.2mm}
System Bandwidth                                               & 10 MHz                              \\
\vspace{0.2mm}
Number of MPCs $L_{b,k}$                                       & ${\cal U}(L_{b,k};40, 100)$                      \\
\vspace{0.2mm}
Path delay of the $l$-th MPC  $\tau_{b,k}^l$                   & ${\cal U}(\tau_{b,k}^l;0, N_{CP}/B_s)$                \\
\vspace{0.2mm}
Angular spread in degree                                       & $10^{\circ}$                        \\
\vspace{0.2mm}
Number of SS-GAMP iteration $T_{\rm amp}$                      & 20                                 \\
\vspace{0.2mm}
Number of turbo iteration $T_{\rm tur}$                        & 10                                 \\
\vspace{0.2mm}
Termination threshold $\eta$                                   & $10^{-5}$                           \\
\vspace{0.2mm}
Path loss $\rho_{b,k}$ at distance $d_{b,k}$ in km & $128.1+37.6{\rm log}_{10}(d_{b,k})$ \\ \bottomrule
\end{tabular}
\vspace{-2.5mm}
\end{table}

For performance evaluation, we consider the detection error probability of AUD $P_e$ and the normalized mean squire error (NMSE) of CE, which are respectively defined as follows
\begin{align}
P_e &= \frac{\sum\nolimits_k \left|{\widehat \alpha_k} - \alpha_k\right|} K,\\
{\rm NMSE} &= 10{\rm log}_{10}\frac{\sum\nolimits_p\sum\nolimits_k \left\|{\widehat {\bf h}_{p,b^*,k}} - {\bf h}_{p,b^*,k}\right\|_2^2} {\sum\nolimits_p\sum\nolimits_k \left\|{\bf h}_{p,b^*,k}\right\|_2^2}.
\end{align}
For AUD in the edge computing paradigm, we obtain the activity estimate of the $k$-th UE ${\widehat \alpha_k}$ based on the signals received at its nearest one DPU-AP and $(N_{co}-1)$ APs.
Moreover, due to the smallest path loss, the UE is expected to be served by the nearest AP in data transmission phase.
Thus, for CE, we mainly focus on the estimation reliability of the channel between the $k$-th UE and the $b^*$-th AP, which has the shortest spatial distance with the $k$-th UE.
We compare the proposed schemes with the following benchmarks:

\begin{itemize}
\item{\textbf{Baseline 1} (Multi-cell non-cooperative massive MIMO-based IoT): To verify the superiority of the proposed cell-free massive MIMO-based IoT architecture, a conventional multi-cell non-cooperative massive MIMO-based IoT architecture is compared as the baseline~1, where each AP (i.e., massive MIMO BS) only serves its own cell's UEs without multi-cell cooperation and treats the inter-cell interference as noise~\cite{Chen_TWC'19}.}

\item{\textbf{Baseline 2} (SS-GAMP-based joint AUD and CE): To show the effectiveness of the proposed SIC-based AUD and CE algorithm, the conventional spatial domain-based massive access scheme is compared as the baseline~2, where the proposed SS-GAMP algorithm is applied to~(\ref{Eq:Cloud_Process_Spa}) for joint AUD and CE.}

\item{\textbf{Baseline 3} (SS-GAMP algorithm using SLM to process quantized signals): To demonstrate the advantage of the proposed SS-GAMP-based joint AUD and CE scheme as well as the SIC-based scheme in the case of processed signals with low-resolution quantization, we compare those two schemes based on SS-GAMP algorithm only using SLM as baseline~3.}
\end{itemize}

\subsection{Superiority of Cell-Free Massive MIMO}
\label{Sec:VI-A}

\begin{figure}[t]
    \captionsetup{font={footnotesize}, name = {Fig.}, labelsep = period}
    \centering
    \includegraphics[width=0.73\columnwidth,keepaspectratio]
    {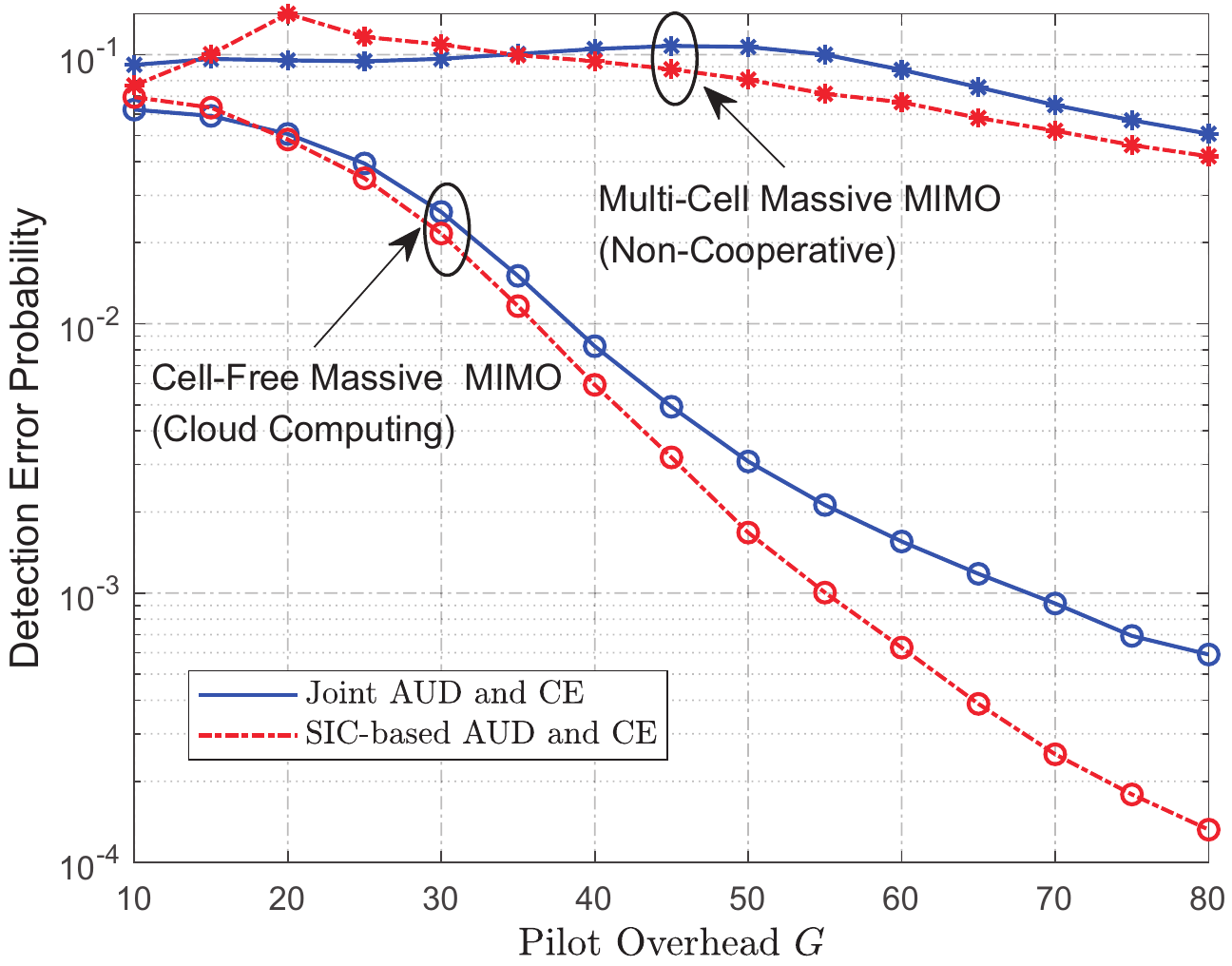}
	\caption{AUD performance comparison of the proposed cloud computing-based scheme and \emph{Baselines 1 and 2}, where $M_c = 16$, ${\widetilde P} =1$, and $T_{\rm sic} = 3$.}
    \label{Fig:7}
\end{figure}

\begin{figure*}[t]
    \captionsetup{font={footnotesize}, name = {Fig.}, labelsep = period}
    \begin{minipage}[t]{1\columnwidth}\centering\includegraphics[width=2.5in]{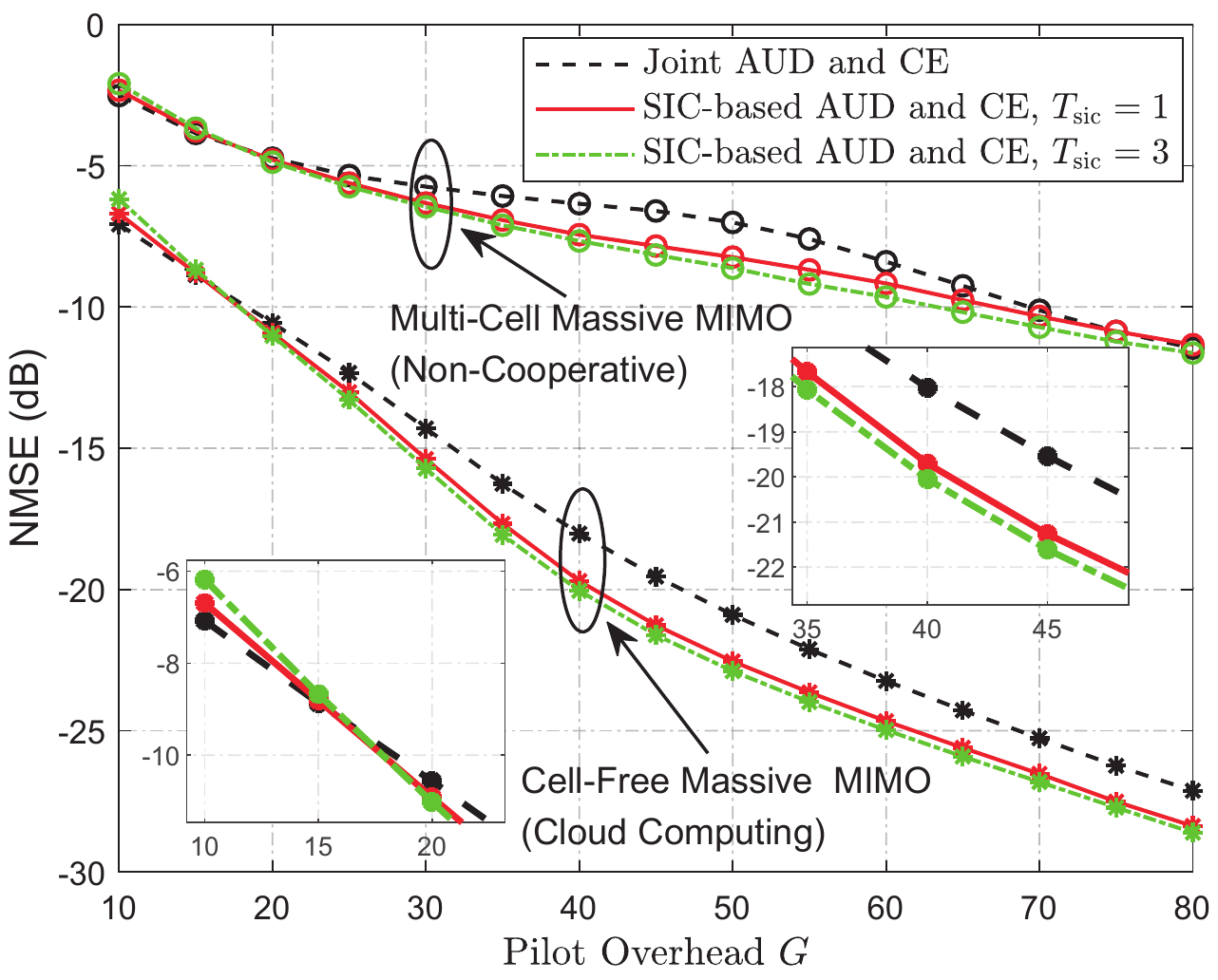}
    \caption{CE performance comparison of the proposed cloud computing-based scheme and \emph{Baselines 1 and 2}, where $M_c = 16$, ${\widetilde P} =1$, and $T_{\rm sic} = 3$.}
    \label{Fig:8}
    \end{minipage}\hfill
    \begin{minipage}[t]{1\columnwidth}\centering\includegraphics[width=2.5in]{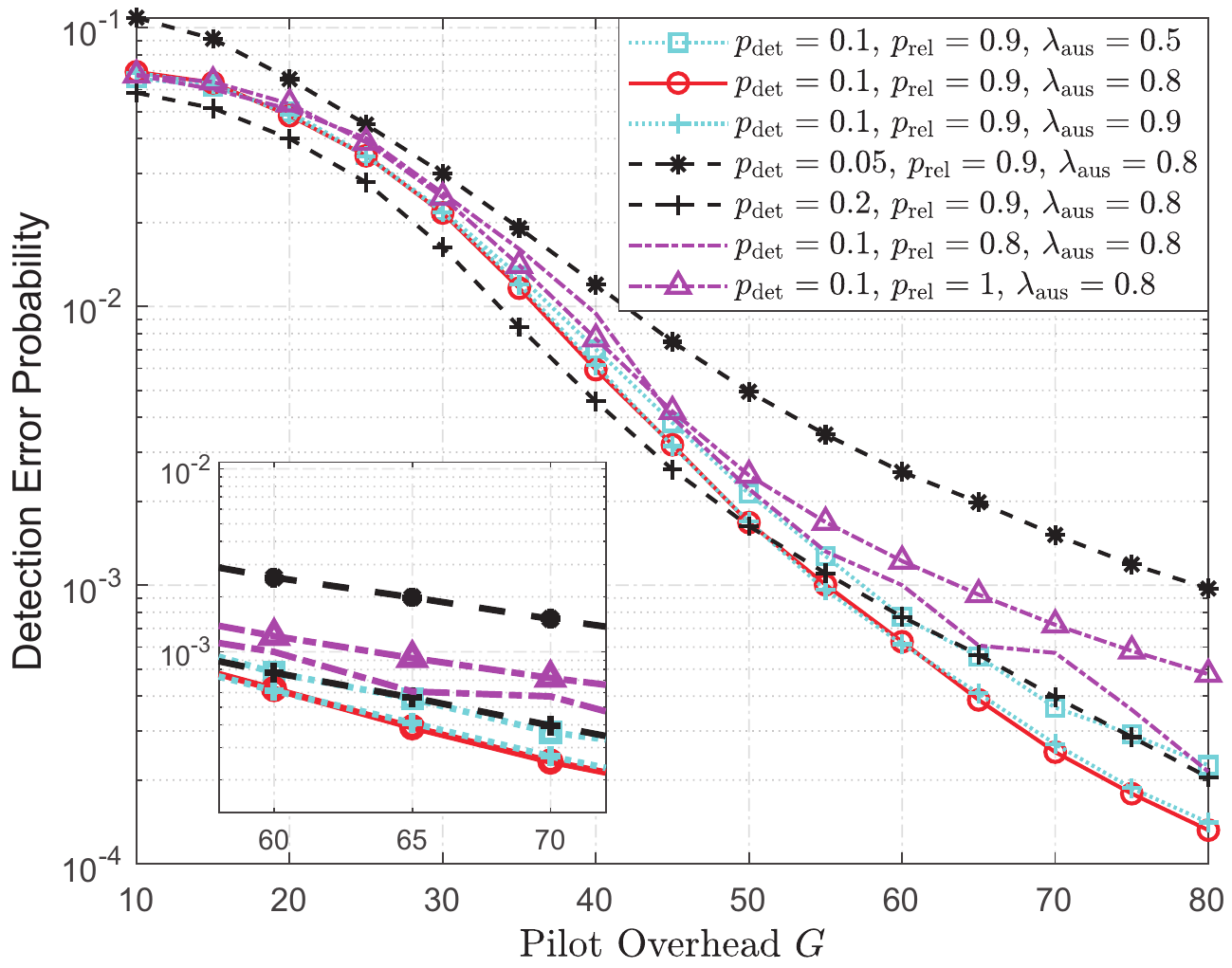}
    \caption{AUD performance of the proposed cloud computing-based scheme under different $\lambda_{\rm aus}$, $p_{\rm det}$, and $p_{\rm rel}$, where $M_c = 16$, ${\widetilde P} =1$, and $T_{\rm sic} = 3$.}
    \label{Fig:9}
    \end{minipage}
\end{figure*}

\begin{figure*}[t]
    \captionsetup{font={footnotesize}, name = {Fig.}, labelsep = period}
    \begin{minipage}[t]{1\columnwidth}\centering\includegraphics[width=2.5in]{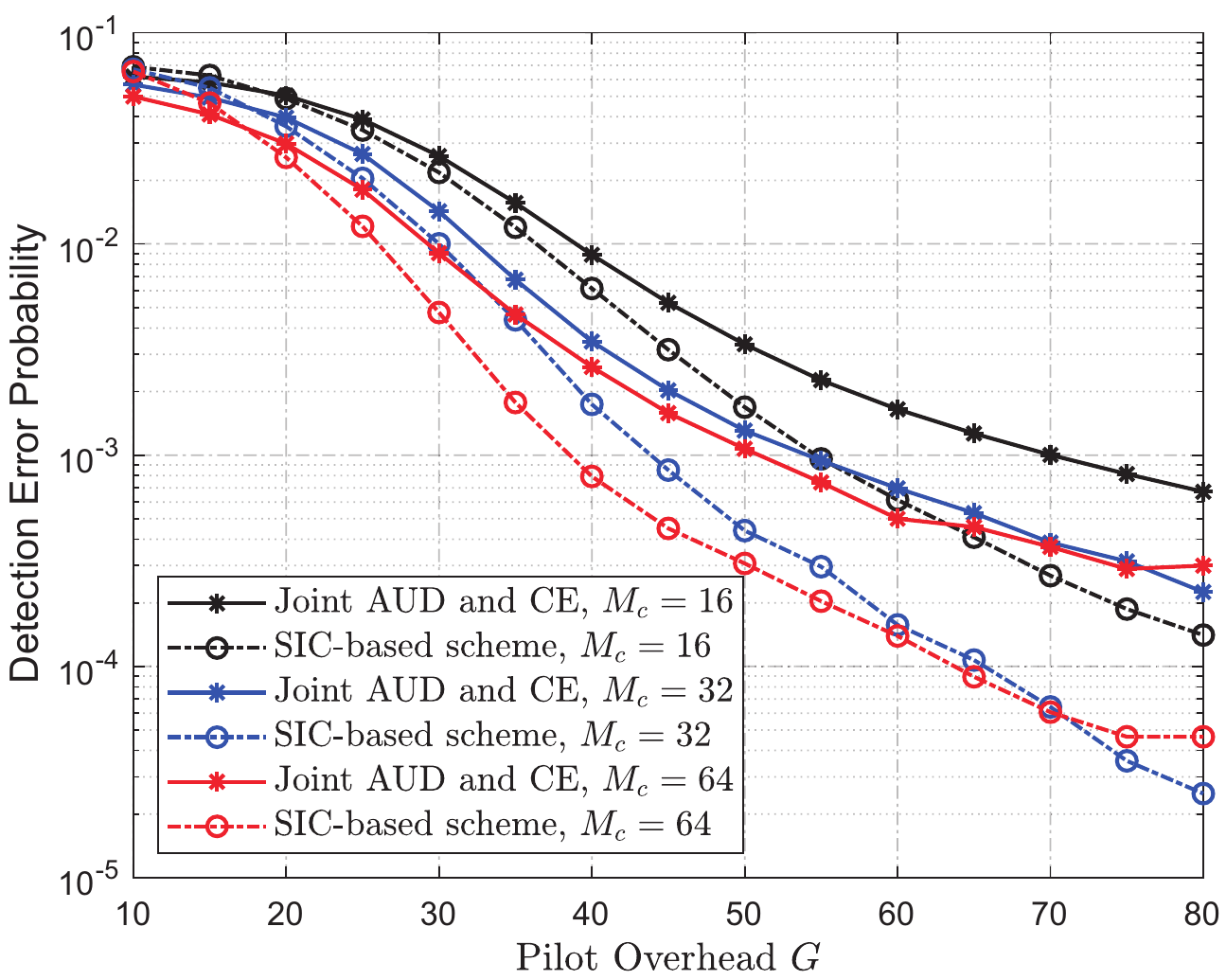}
    \caption{AUD performance of the proposed SIC-based scheme and \emph{Baseline~2} for different AP antennas $M_c$, where the cloud computing is considered.}
    \label{Fig:10}
    \end{minipage}\hfill
    \begin{minipage}[t]{1\columnwidth}\centering\includegraphics[width=2.5in]{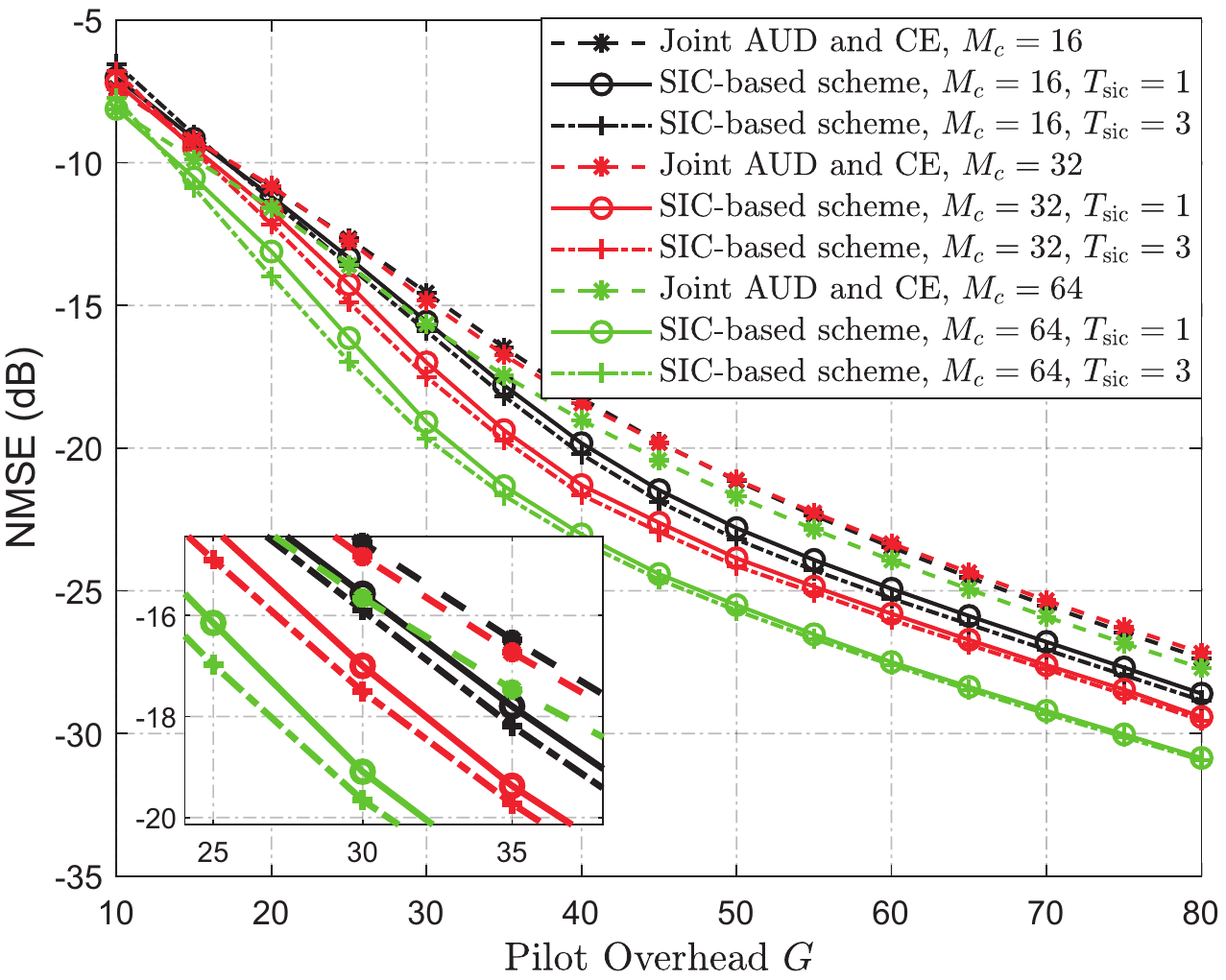}
    \caption{CE performance of the proposed SIC-based scheme and \emph{Baseline 2} for different AP antennas $M_c$, where the cloud computing is considered.}
    \label{Fig:11}
    \end{minipage}
\end{figure*}

This section validates the superiority of the proposed cell-free massive MIMO-based IoT architecture, where $Q = 10$ is considered.
Fig.~\ref{Fig:7} compares the AUD performance of the proposed cloud computing-based scheme and \emph{Baselines~1 and~2}.
It can be observed that the cloud computing-based processing paradigm proposed in cell-free massive MIMO-based IoT can achieve a much better AUD performance than multi-cell non-cooperative massive MIMO-based IoT.
The reason is that there are no cell boundaries in cell-free massive MIMO systems, and the inter-cell interference can be avoided via the APs' cooperation.
Moreover, for $G \ge 20$, by further leveraging the angular-domain sparsity and the idea of SIC, the proposed SIC-based AUD and CE algorithm outperforms the conventional joint AUD and CE scheme, which is only based on the spatial-domain channel model.
However, for the very low pilot overhead region (e.g., $G < 20$), the joint AUD and CE scheme performs better than the proposed SIC-based method.
This is because the AUS and CSI estimates are extremely inaccurate in this case, which leads to the error propagation of SIC.
Fig.~\ref{Fig:8} depicts the CE performance of the considered schemes, which further validates the superiority of the proposed cell-free massive MIMO architecture and the SIC-based AUD and CE scheme for massive access.
Here, the partially enlarged views show the NMSE performance for the pilot overhead regions $G \in \left[10, 20\right]$ and $G \in \left[30, 45\right]$, respectively.
Fig.~\ref{Fig:9} further studies the influence of parameters $\lambda_{\rm aus}$, $p_{\rm det}$, and $p_{\rm rel}$ on massive access performance.
As can be observed, when $\lambda_{\rm aus} = 0.8$, $p_{\rm det} = 0.1$, and $p_{\rm rel}= 0.9$, the proposed approach achieves the best AUD performance.

%\begin{figure}[t]
%    \captionsetup{font={footnotesize}, name = {Fig.}, labelsep = period}
%    \centering
%    \includegraphics[width=0.73\columnwidth,keepaspectratio]
%    {Fig8/Fig_8.eps}
%	\caption{CE performance comparison of the proposed cloud computing-based scheme and \emph{Baselines 1 and 2}, where $M_c = 16$, ${\widetilde P} =1$, and $T_{\rm sic} = 3$.}
%    \label{Fig:8}
%    \vspace{-0.5mm}
%\end{figure}
%
%\begin{figure}[t]
%    \captionsetup{font={footnotesize}, name = {Fig.}, labelsep = period}
%    \centering
%    \includegraphics[width=0.73\columnwidth,keepaspectratio]
%    {Fig9/Fig_9.eps}
%	\caption{AUD performance of the proposed cloud computing-based scheme under different $\lambda_{\rm aus}$, $p_{\rm det}$, and $p_{\rm rel}$, where $M_c = 16$, ${\widetilde P} =1$, and $T_{\rm sic} = 3$.}
%    \label{Fig:9}
%\end{figure}

\begin{figure*}[t]
    \captionsetup{font={footnotesize}, name = {Fig.}, labelsep = period}
    \begin{minipage}[t]{1\columnwidth}\centering\includegraphics[width=2.5in]{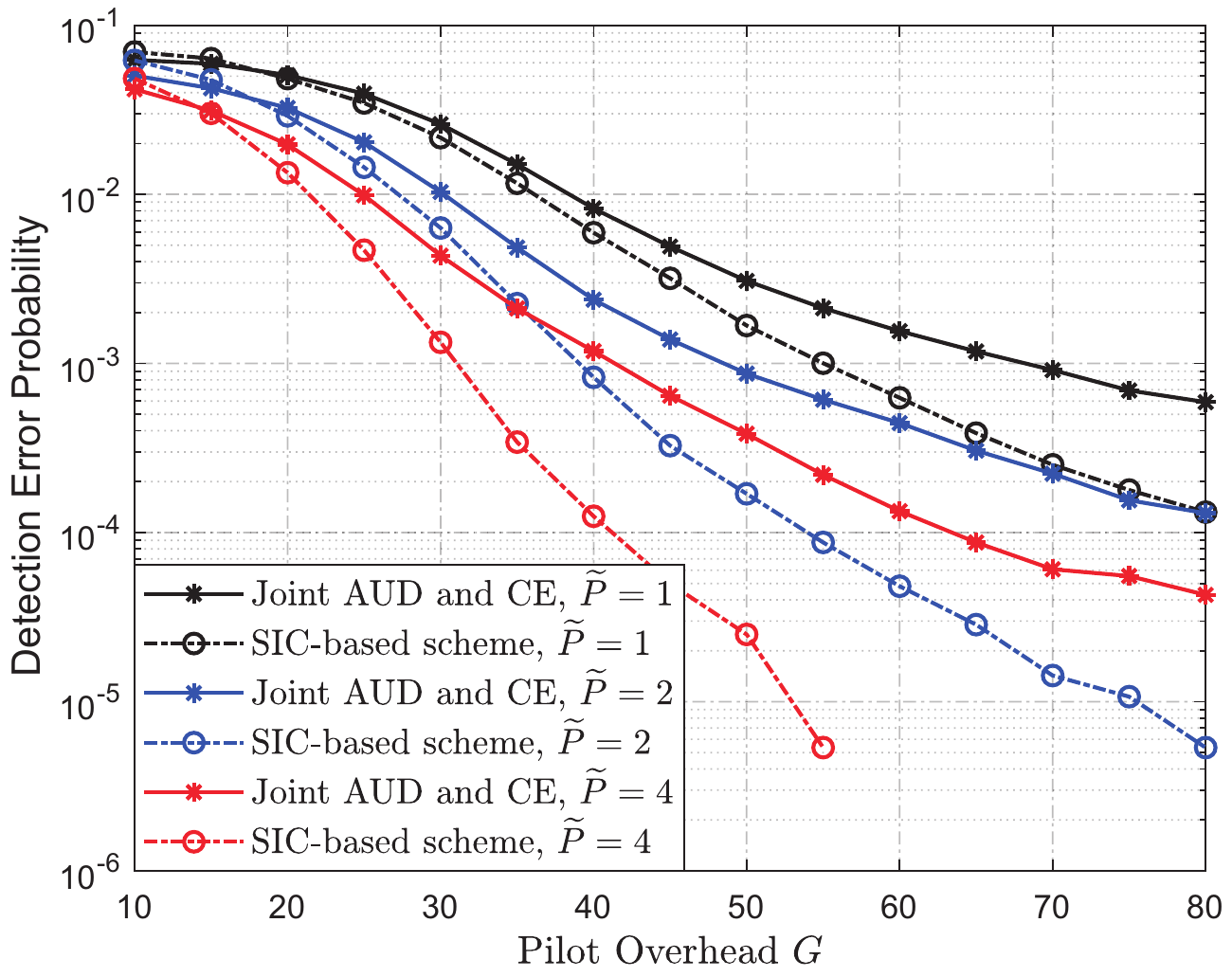}
    \caption{AUD performance of the proposed SIC-based scheme and \emph{Baseline 2} for different ${\widetilde P}$, where the cloud computing and $M_c = 16$ are considered.}
    \label{Fig:12}
    \end{minipage}\hfill
    \begin{minipage}[t]{1\columnwidth}\centering\includegraphics[width=2.5in]{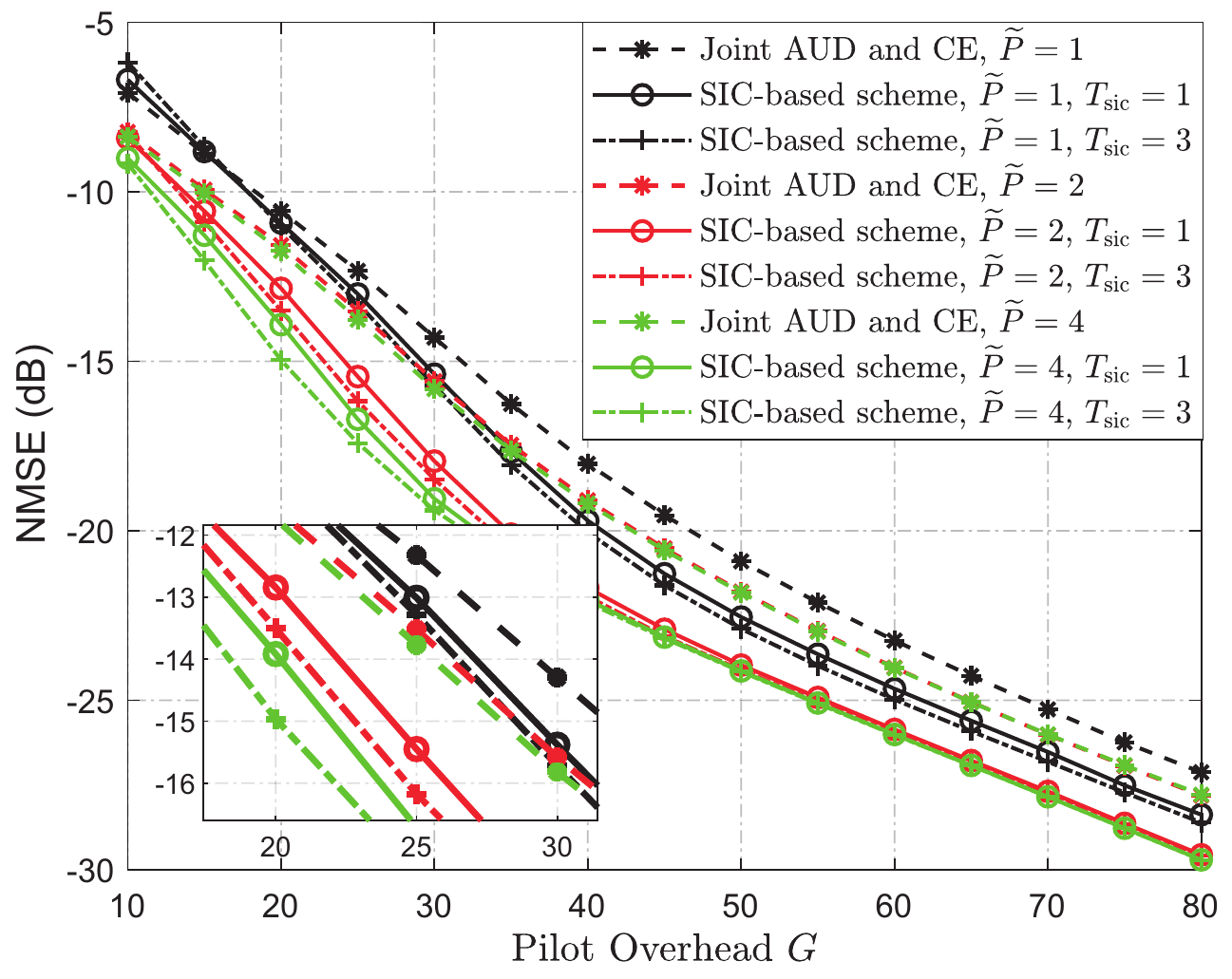}
    \caption{CE performance of the proposed SIC-based scheme and \emph{Baseline 2} for different ${\widetilde P}$, where the cloud computing and $M_c = 16$ are considered.}
    \label{Fig:13}
    \end{minipage}
\end{figure*}

Fig.~\ref{Fig:10} and Fig.~\ref{Fig:11} verify the superiority of massive MIMO-based APs for grant-free massive access, where $T_{\rm sic} = 3$ and ${\widetilde P} =1$ are considered.
It is clear that the proposed cloud computing-based scheme can achieve a better performance by equipping more antennas at the APs.
For AUD based on the spatial-domain channel model, a larger number of AP antennas enhances the spatial-domain structured sparsity of the channel matrix $\{  {\bf H}_{p} \}_{p=1}^P$, which improves the accuracy of AUS estimate.
On the other hand, a massive number of antennas can promote the angular-domain sparsity of massive MIMO channels, which can be leveraged to improve the CE performance.
If the APs have a relatively small number of antennas (e.g., $M_c = 16$), the angular-domain sparsity of the massive MIMO channels would be weaken, and the performance of the proposed scheme would be degraded.
Hence, the proposed scheme shows its superiority for massive MIMO cases.
Fig.~\ref{Fig:12} and Fig.~\ref{Fig:13} show that the increased ${\widetilde P}$ also improves the AUD and CE performance of the proposed cloud computing-based scheme, where $T_{\rm sic} = 3$ is considered.
This is because a larger ${\widetilde P}$ can also enhance the spatial-domain structured sparsity of the channel matrix $\{  {\bf H}_{p} \}_{p=1}^P$.

%\begin{figure}[t]
%    \captionsetup{font={footnotesize}, name = {Fig.}, labelsep = period}
%    \centering
%    \includegraphics[width=0.73\columnwidth,keepaspectratio]
%     {Fig10/Fig_10.eps}
%	\caption{AUD performance of the proposed SIC-based scheme and \emph{Baseline~2} for different AP antennas $M_c$, where the cloud computing is considered.}
%    \label{Fig:10}
%\end{figure}
%
%\begin{figure}[t]
%    \captionsetup{font={footnotesize}, name = {Fig.}, labelsep = period}
%    \centering
%    \includegraphics[width=0.73\columnwidth,keepaspectratio]
%    {Fig11/Fig_11.eps}
%	\caption{CE performance of the proposed SIC-based scheme and \emph{Baseline 2} for different AP antennas $M_c$, where the cloud computing is considered.}
%    \label{Fig:11}
%\end{figure}

\subsection{Comparison of Cloud Computing and Edge Computing Paradigms}
\label{Sec:VI-B}

\begin{figure*}[t]
    \captionsetup{font={footnotesize}, name = {Fig.}, labelsep = period}
    \centering
    \subfigure[]{\includegraphics[width=0.73\columnwidth]{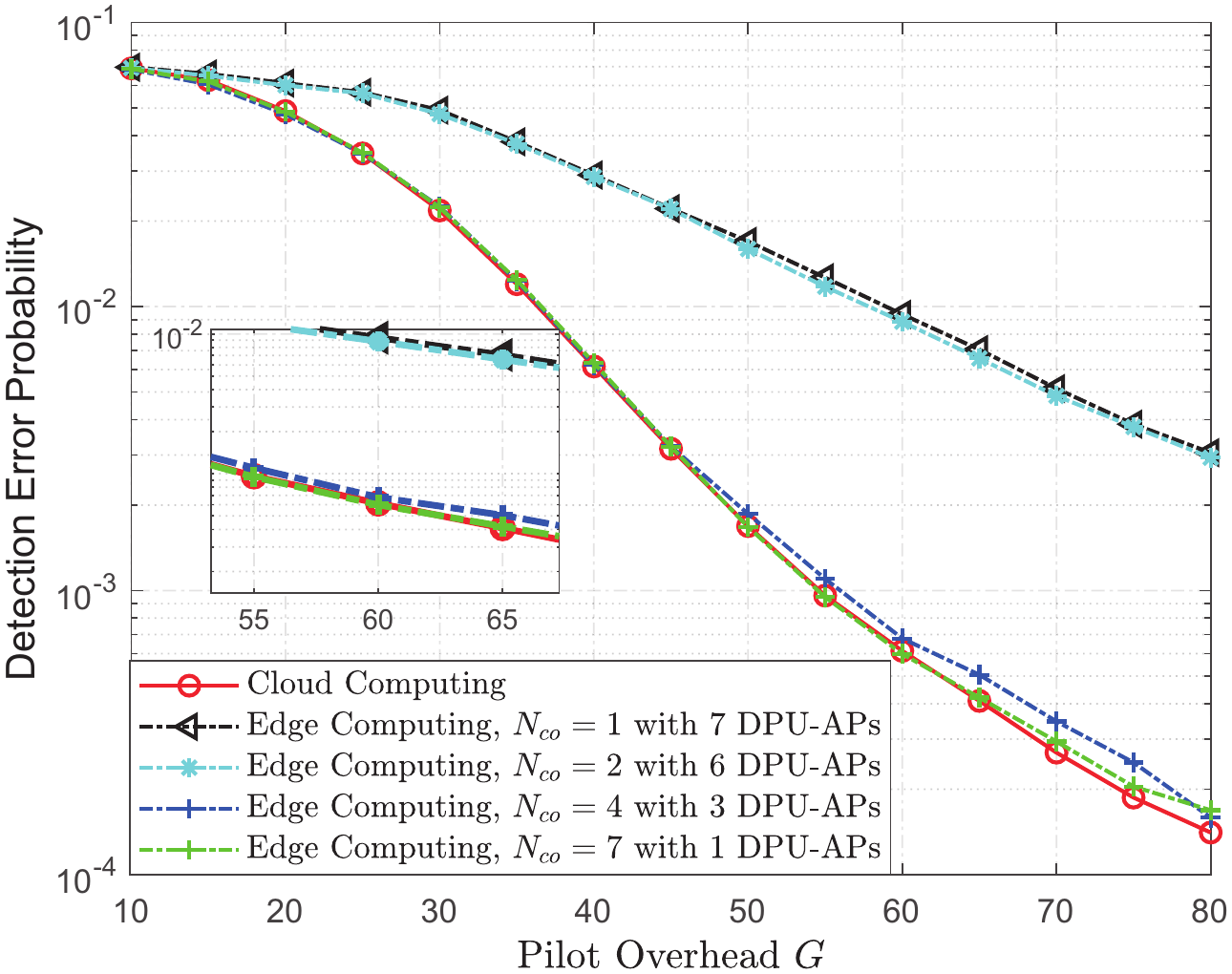}}\hspace{27mm}
    \subfigure[]{\includegraphics[width=0.73\columnwidth]{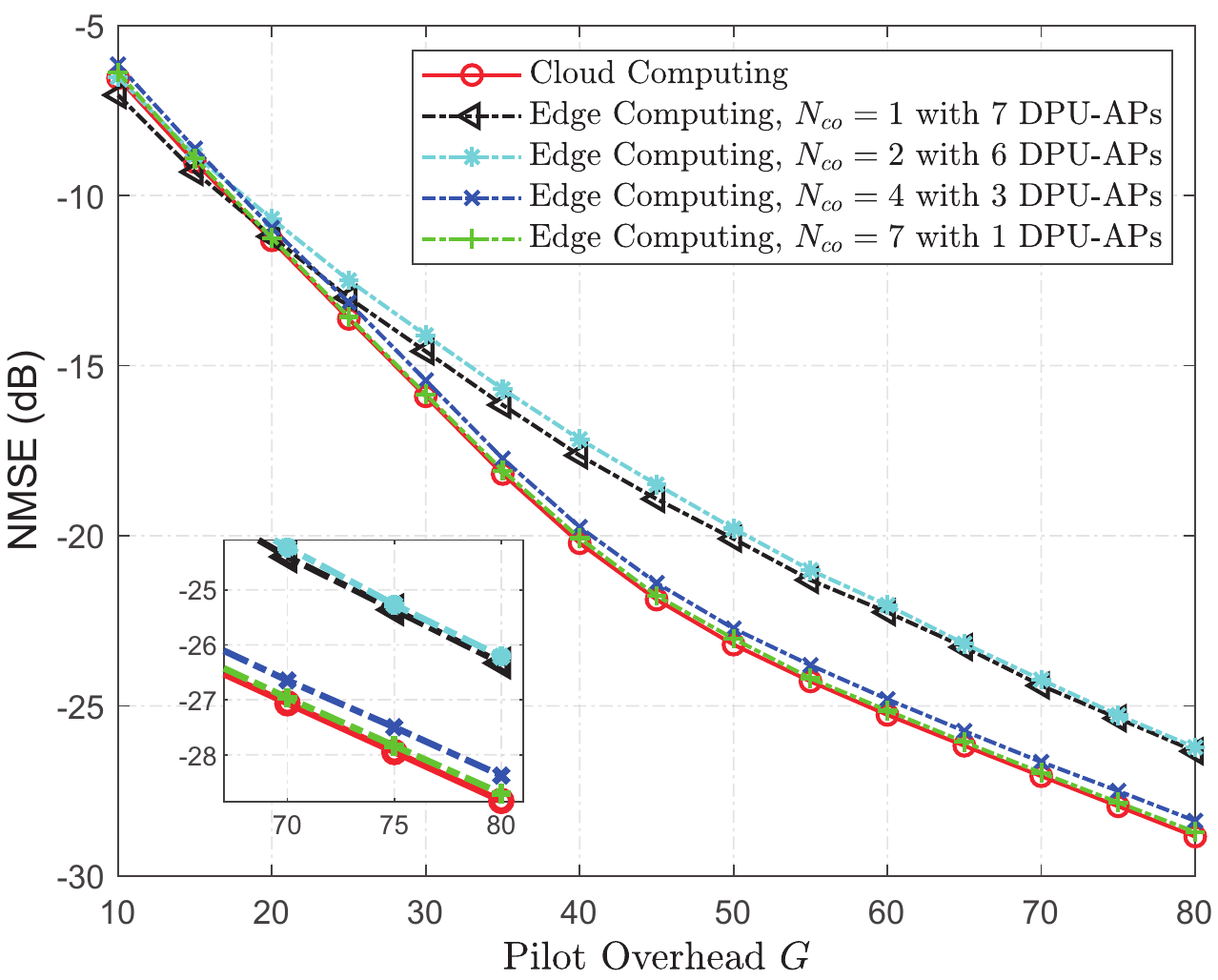}}
    \caption{Comparison of the proposed cloud computing and edge computing paradigms, where the proposed SIC-based scheme ($T_{\rm sic} = 3$) is considered, $M_c = 16$, ${\widetilde P} = 1$, and $Q = 10$: (a) AUD performance; (b) CE performance. For edge computing paradigm, the number of required DPU-APs is provided for different $N_{co}$.}
    \label{Fig:14}
\end{figure*}

For the proposed SIC-based massive access scheme designed for cell-free massive MIMO-based IoT, we further compare two computing paradigms for the processing of AUD and CE, as shown in Fig.~\ref{Fig:14}.
By increasing the number of APs for cooperation, i.e., $N_{co}$, the AUD and CE performance of edge computing approaches that of cloud computing.
Furthermore, we observe that only $N_{co} = 4$ APs are required for edge computing to obtain almost the same performance of cloud computing.
This is because the channel gains from a specific active UE to the far away APs are approximate zero, the signals received at the remote APs can not further improve the AUD and CE performance.
When all the $N$ APs, consisting of a DPU-AP and $\left(N-1\right)$ conventional APs, cooperate, i.e., $N_{co} = N$, the edge computing paradigm is equivalent to the cloud one.
Meanwhile, note that there is a tradeoff between the performance and the cost of practical DPU-AP deployment.
Compared to the cloud computing, the edge computing can reap a more cost-effective cooperation (i.e., CPU burden, backhaul cost, and response time), while may increase the price of network deployment, i.e., the DPU-APs should employ DPUs.

\subsection{Massive Access Under Limited Backhaul Capacity}
\label{Sec:VI-C}

\begin{figure*}[t]
    \captionsetup{font={footnotesize}, name = {Fig.}, labelsep = period}
    \begin{minipage}[t]{1\columnwidth}\centering\includegraphics[width=2.5in]{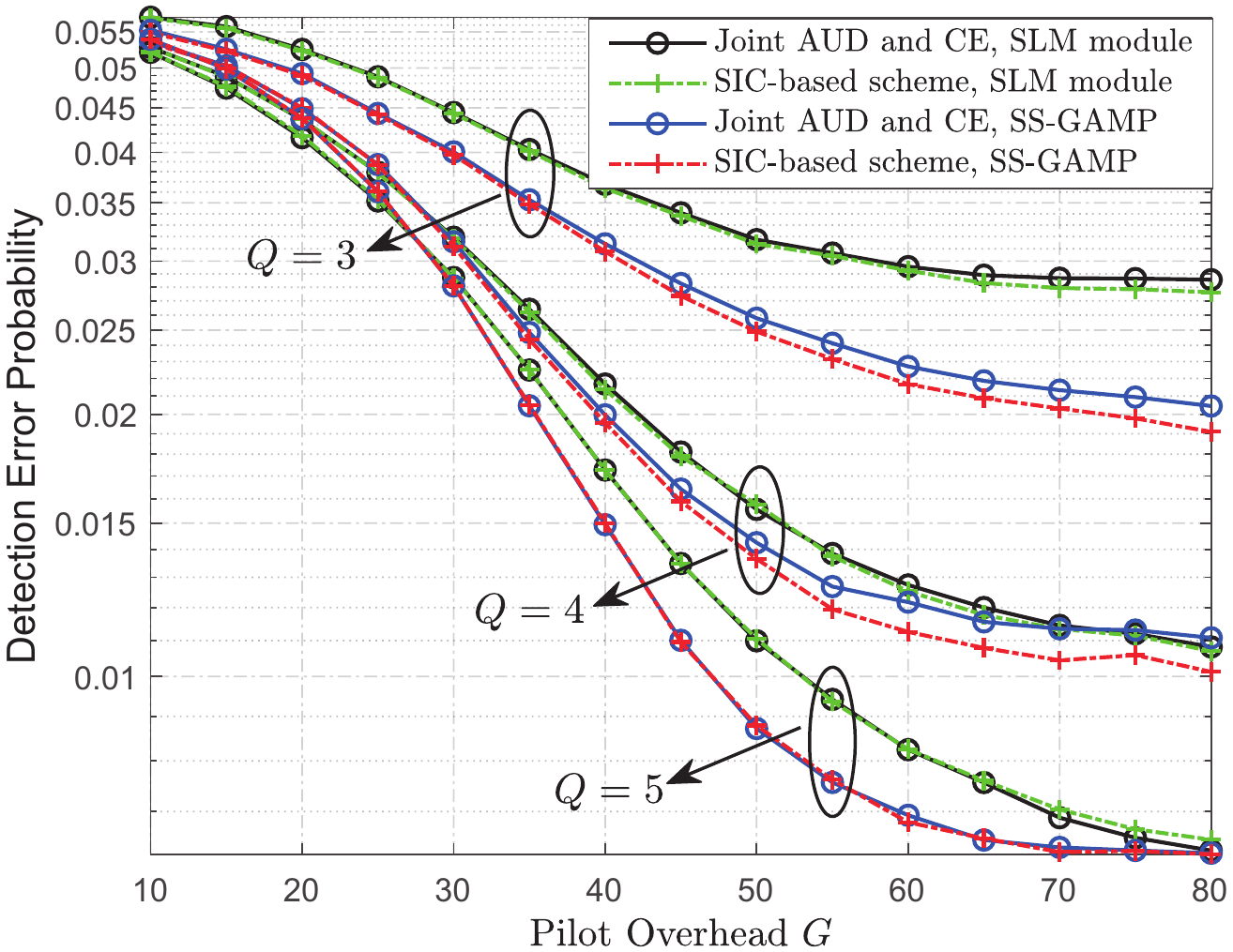}
    \caption{AUD performance of the proposed SS-GAMP-based schemes and \emph{Baseline 3} in low-accuracy quantization cases, where $M_c = 16$, ${\widetilde P} =1$, and $T_{\rm sic} = 3$.}
    \label{Fig:15}
    \end{minipage}\hfill
    \begin{minipage}[t]{1\columnwidth}\centering\includegraphics[width=2.5in]{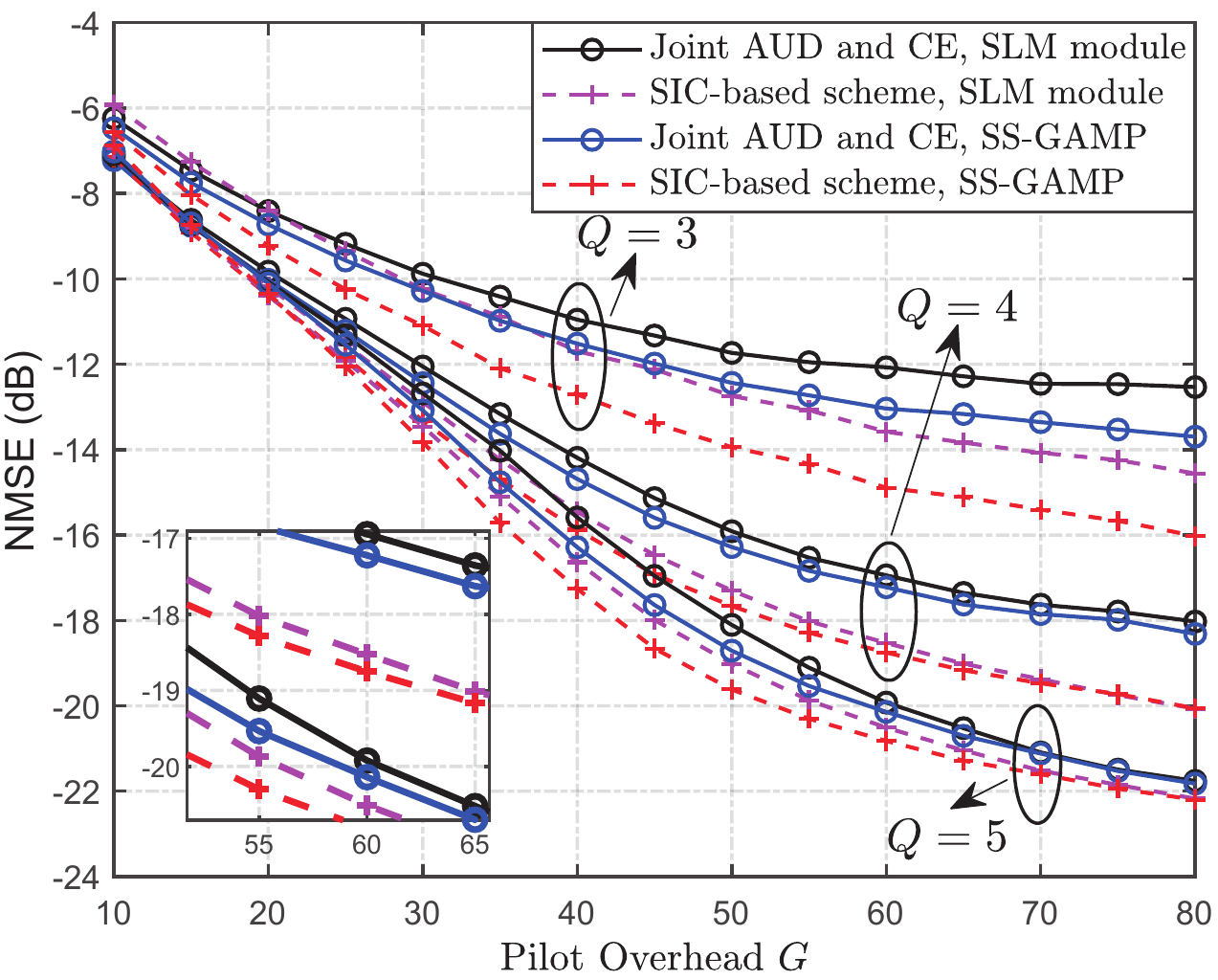}
    \caption{NMSE performance of the proposed SS-GAMP-based schemes and \emph{Baseline 3} in low-accuracy quantization cases, where $M_c = 16$, ${\widetilde P} =1$, and $T_{\rm sic} = 3$.}
    \label{Fig:16}
    \end{minipage}
    \vspace{-1mm}
\end{figure*}

Fig.~\ref{Fig:15} and Fig.~\ref{Fig:16} verify the superiority of the proposed schemes based on SS-GAMP algorithm over those based on baseline 3, where a low-resolution quantization of the processed signals is considered.
Here, the number of quantization bits $Q = 3, 4$, and 5 are investigated.
As can be observed, the proposed SS-GAMP algorithm can achieve a better performance than baseline~3 in both joint AUD and CE scheme and SIC-based scheme.
This is because the quantization is taken into account by using the SS-GAMP algorithm with nonlinear module.

\section{Conclusion}
\label{Sec:VII}

This paper studies grant-free massive access in cell-free massive MIMO-based IoT, where multiple APs cooperate in the network to serve massive UEs.
By exploiting the structured sparsity of the channel matrix, we develop a SS-GAMP algorithm for the CS recovery, where the quantization accuracy of the processed signals is considered.
On this basis, a SIC-based AUD and CE algorithm is further proposed.
Compared to the conventional massive access schemes based on the single-cell or multi-cell non-cooperative network architectures, cell-free massive MIMO can offer better coverage and improve the AUD and CE performance via AP cooperation.
Furthermore, in contrast to the CS algorithms for SLM and the spatial-domain joint AUD and CE scheme, the proposed SIC-based scheme using SS-GAMP algorithm can significantly reduce the access latency in low-resolution quantization cases.
Besides, we consider two computing paradigms, cloud computing and edge computing, to perform AUD and CE.
Numerical simulations suggest that the performance of the edge computing can approach that of cloud computing.
Meanwhile, edge computing makes the cooperation of APs more flexible and alleviates the burden on CPU and backhaul links, while may increase the cost of network deployment.


\ifCLASSOPTIONcaptionsoff
  \newpage
\fi

\end{document}